\documentclass[12pt]{article}
\usepackage{amsthm}
\usepackage{amssymb}
\usepackage{amsmath}
\usepackage{amsfonts}
\usepackage{enumerate}
\usepackage{verbatim}
\usepackage[dvips]{graphicx}
\usepackage[english]{babel}
\usepackage[cp1250]{inputenc}
\usepackage[T1]{fontenc}
\usepackage[a4paper,left=2.85cm,right=2.85cm,top=2.85cm,bottom=2.85cm]{geometry}
\usepackage[numbers]{natbib}

\newtheorem{prop}{Proposition}[section]
\newcommand{\Proof}{\noindent\textbf{{Proof:}}\newline}
\numberwithin{equation}{section}

\newcommand{\cbdu}{\quad\hfill\mbox{$\Box$}\\[3mm]}
\newcommand{\beqo}{\begin{eqnarray*}}
\newcommand{\eeqo}{\end{eqnarray*}\noindent}
\newcommand{\beq}{\begin{eqnarray}}
\newcommand{\eeq}{\end{eqnarray}\noindent}
\def\be{{\mathbb{E}}}
\def\bp{{\mathbb{P}}}
\def\bq{{\mathbb{Q}}}
\def\bqq{{\tilde{\mathbb{Q}}}}
\def\bfi{{\mathcal{F}}}

\begin{document}

\title{\textbf{BSDEs with time-delayed generators of a moving average type with applications to non-monotone preferences}\footnote{Acknowledgements: The research is supported by the Foundation for Polish Science. }}
%The author would like to thank two anonymous referees for pointing out deficiencies and the Editors for an encouragement to improve the earlier version of this paper.

\author{\textbf{{\L}ukasz Delong}\\
\footnotesize{Institute of Econometrics, Division of Probabilistic Methods}\\
\footnotesize{Warsaw School of Economics}\\
\footnotesize{Al. Niepodleg{\l}o\'{s}ci 162, 02-554 Warsaw, Poland}\\
\footnotesize{lukasz.delong@sgh.waw.pl}}

\date{}
\maketitle

\newpage
\begin{abstract}
\noindent In this paper we consider backward stochastic differential equations with time-delayed generators of a moving average type. The classical framework with linear generators depending on $(Y(t),Z(t))$ is extended and we investigate linear generators depending on $(\frac{1}{t}\int_0^tY(s)ds, \frac{1}{t}\int_0^tZ(s)ds)$. We derive explicit solutions to the corresponding time-delayed BSDEs and we investigate in detail main properties of the solutions. An economic motivation for dealing with the BSDEs with the time-delayed generators of the moving average type is given. We argue that such equations may arise when we face the problem of dynamic modelling of non-monotone preferences. We model a disappointment effect under which the present pay-off is compared with the past expectations and a volatility aversion which causes the present pay-off to be penalized by the past exposures to the volatility risk.
\noindent \textbf{Running title:} BSDEs with time-delayed generators of a moving average type.
\end{abstract}
\newpage

\section{Introduction}
A backward stochastic differential equation (BSDE) driven by a Brownian motion $W$ is an equation of the following form
\beq\label{bsde}
Y(t)=\xi-\int_t^Tg(s,Y(s),Z(s))ds-\int_t^TZ(s)dW(s),\quad 0\leq t\leq T,
\eeq
where $\xi$ is a terminal condition and $g$ is a generator of the equation. It is now known that BSDEs \eqref{bsde} could be used to define dynamic pricing principles, dynamic risk measures and recursive utilities. This modelling approach has been initiated in Duffie \& Epstein \cite{duffie}, where recursive utilities as solutions to BSDEs are derived, and in Rosazza Gianin \cite{gianin}, where dynamic risk measures as $g$-expectations are introduced.\\
\indent Barrieu \& El Karoui \cite{barrieu}, El Karoui \& Ravenelli \cite{ravanelli}, Lazrak \& Quenez \cite{lazrak}, among others, strongly advocate the construction of new pricing principles, risk measures and utilities via an interpretable generator $g$ and the corresponding BSDE. Following the authors, let us interpret the process $Y$ as a price or a utility. Heuristically, the BSDE \eqref{bsde} implies the relation
\beq\label{intery}
\be[dY(t)|\bfi_t]=g(t,Y(t),Z(t))dt,\quad 0\leq t< T.
\eeq
Motivated by \eqref{intery} we can now interpret $g$ as a local preference-based rule which describes subjective beliefs concerning the expected change in the price or the utility. We have another heuristic relation for variance of the infinitesimal change
\beq\label{interz}
\mathbb{V}[dY(t)|\bfi_t]=Z^{2}(t)dt,\quad 0\leq t< T,
\eeq
and motivated by \eqref{interz} we can interpret $Z$ as an intensity of variability or volatility in the price or the utility. With these interpretations of $g, Y, Z$, we can first define a local pricing rule or an instantaneous utility via the coefficient $g$ depending on $(Y,Z)$. The coefficient $g$ quantifies risk preferences and beliefs, reflecting how the expected change in the price or the utility is related to the price or utility and their variability. Given the local valuation rule and the local risk aversion coefficient $g$ we can solve the BSDE \eqref{bsde} with the generator $g$ to obtain the global valuation rule $Y$. \\
\indent In the literature on risk and utility modelling via BSDEs it is assumed that the generator $g$ has a Markovian structure with respect to the pair $(Y,Z)$ in the sense that the generator $g$ at time $t$ depends only on the values of $Y(t), Z(t)$ at that time $t$. Such Markovian structure is enforced by the current state of the theory on backward stochastic differential equations. However, a Markovian structure seems to be too restrictive when investors' preferences are concerned. It is unlikely that investors form their views and choices (quantified via the value of $g$ at $t$) only based on the current information without taking into account the history. It seems more reasonable to assume that investors have the memory, compare previous and current opportunities, take into account experienced trends in the prices or satisfaction from the past consumptions, form a priori expectations about the projects, compare their past expectations with the current pay-offs, study the risk factors and the realized gains, and finally make decisions. Such behaviours cannot be obtained with Markovian generators in \eqref{bsde}.\\
\indent The memory of the past events clearly matters. The memory is mentioned in Loewenstein \cite{loewen} as one of three factors which could help in understanding preferences and intertemporal choices of agents. Hence, a motivation for considering non-Markovian generators $g$ in \eqref{bsde} and for a study of such equations arises. This leads us to BSDEs with time-delayed generators. A backward stochastic differential equation with a time-delayed generator driven by a Brownian motion $W$ is an equation
\beq\label{tdbsde}
Y(t)=\xi-\int_t^Tg(s,Y_s,Z_s)ds-\int_t^TZ(s)dW(s),\quad 0\leq t\leq T,
\eeq
where the generator $g$ at time $s$ is allowed to depend on the past values of a solution $Y_s:=(Y(u))_{0\leq u\leq s}, Z_s:=(Z(u))_{0\leq u\leq s}$ up to time $s$. This new class of BSDEs has been introduced in Delong \& Imkeller \cite{delong1} and Delong \& Imkeller \cite{delong2}. Notice that by a construction of the generator, the BSDE \eqref{tdbsde} incorporates the memory into the system. We believe that BSDEs with time-delayed generators could prove to be useful in modelling a feedback of the past experiences into the future expectations. In particular, we claim that BSDEs with time-delayed generators have a potential in modelling non-monotone preferences under a disappointment effect and a volatility aversion. \\
\indent The classical utility theory assumes that an increase in a consumption in any period increases the total utility. This effect of monotonicity is modelled in the framework of BSDEs \eqref{bsde} as a solution to a BSDE satisfies a comparison principle. However, there exists doubts in economics whether any increase in consumption should really increase the utility. Rozen \cite{rozen} lays down the theory of non-monotone preferences. The author observes that "individual total welfare might not increase if a period of luxury life is followed by an individual's return to a humble lifestyle". Rozen \cite{rozen} argues that a temporary increase in a consumption does not have to lead to an increase in the utility and increasing only a finite number of elements in a consumption stream may dampen future enjoyment of that stream. A short-term increase in a consumption is not beneficial if a period of an upturn has ended and is followed by a period of a downturn. An increase in the utility takes place only when an increase in a consumption in permanent. As already mentioned, the classical BSDEs \eqref{bsde} cannot model such non-monotone preferences. However, it is shown in Delong \& Imkeller \cite{delong1} that BSDEs with time-delayed generators \eqref{tdbsde} may not satisfy a comparison principle. Hence, these new time-delayed equations might be the right framework to capture non-monotone preferences. Rozen \cite{rozen} relates a non-monotonicity in the preferences to a disappointment effect. A disappointment is a concept which has strong foundations in economics. The paper by Bell \cite{bell} starts with the question: "are you really pleased when your boss has given you 5000\$ bonus if you were expecting 10000\$?" and the answer depends on how the past expectations are related to the outcomes. A disappointment arises in a situation when the a priori expectation does not meet the real outcome and the bonus of 5000\$ could be assessed as a loss relative to the expectation of 10000\$. Risk preferences should be different for an investor who expects to be rich but achieves a modest success than for an investor who starts out with with low expectations and achieves the same modest success, see also Dybvig \& Rogers \cite{rogers} and their motivation for considering a disappointment effect. Clearly, a disappointment is related to our memory and depends on a feedback of the past experiences into the future expectations. Such a feedback could only be modelled by a time-delayed generator. Interestingly, Loewenstein \& Prelec \cite{loewen2} notice that non-monotone preferences could also be induced by an aversion against volatility.\\
\indent We advocate the use of BSDEs with time-delayed generators in dynamic modelling of non-monotone preferences under a disappointment effect and a volatility aversion. Under the disappointment effect the present pay-off is compared with the past expectations and the volatility aversion causes the present pay-off to be penalized by the past exposures to the volatility risk. We investigate the following equations
\beq\label{mybsde}
Y(t)&=&\xi-\beta\int_t^T\frac{1}{s}\int_0^sY(u)duds-\int_t^TZ(s)dW(s),\quad 0\leq t\leq T,\nonumber\\
Y(t)&=&\xi-\beta\int_t^T\frac{1}{s}\int_0^sZ(u)duds-\int_t^TZ(s)dW(s),\quad 0\leq t\leq T,
\eeq
and in the dynamic setting we succeed in capturing behaviours which are pointed out by Bell \cite{bell}, Rozen \cite{rozen} and Loewenstein \& Prelec \cite{loewen2}. To the best of our knowledge this paper provides a first attempt to use BSDEs in modelling non-monotone preferences.\\
\indent This paper contributes to the theory of BSDEs by considering a very important class of linear BSDEs with time-delayed generators \eqref{mybsde}. The above introduction motivates a detailed mathematical study of our equations. We remark that the economic literature indicates that a disappointment effect and an aversion against volatility should be taken into account in the study of decision making under uncertainty. Hence, the properties of our new dynamic models should be carefully investigated. From the mathematical point of view linear equations are always fundamental, their understanding is crucial and they serve as a starting point for more advanced analysis. We point out that the considered generators in \eqref{mybsde}, in spite of linearity, do not satisfy the Lipschitz continuity condition
\beqo
|g(t,y_y,z_t)-g(t,\tilde{y}_t,\tilde{z}_t)|\leq L\big(\int_0^t|y(s)-\tilde{y}(s)|ds+\int_0^t|z(s)-\tilde{z}(s)|ds\big),
\eeqo
which is assumed in Delong \& Imkeller \cite{delong1}, Delong \& Imkeller \cite{delong2}. In consequence, a new study of \eqref{mybsde} has to be carried out. We derive explicit solutions to the BSDEs \eqref{mybsde} and we investigate in detail some of the main properties of the solutions. This paper supplements recent theoretical results on BSDEs with time-delayed generators from Delong \& Imkeller \cite{delong1}, Delong \& Imkeller \cite{delong2} and Dos Reis et al \cite{gonzalo}. We believe that our results are beneficial for the developing theory of the new type of BSDEs.\\
\indent This paper is structured as follows. In Section 2 we motivate the form of the generators. In Section 3 we investigate the BSDE based on the moving average of $Y$ and in Section 4 we deal with the BSDE based on the moving average of $Z$. Economic implications are discussed in Section 5. The Appendix recalls some results on the modified Bessel functions of the first and the second kind which we use in Section 3.

\section{The form of the driving generator}
\indent The classical utility model assumes that the utility $Y$ from the future consumption stream $c$ can be valued as
\beqo
Y(t)=\be\Big[\int_t^Tu(c(s))ds\big| \bfi_t\Big],\quad 0\leq t\leq T,
\eeqo
where $u$ denotes an instantaneous utility. This simple model arises when the driving generator in \eqref{bsde} takes the form
\beq\label{utilitygen}
g(t,Y(t),Z(t))=-u(c(t)).
\eeq
Detemple \& Zapatero \cite{detemple1} and Detemple \& Zapatero \cite{detemple} postulate to add a habit process $z(t;c_t)$ into the utility. The authors motivate that the instantaneous utility or the infinitesimal expected change in the utility process $Y$ depends not only on the current consumption as in \eqref{utilitygen} but also is influenced by satisfaction from the past consumption. It is economically reasonable to assume that the investor values the utility from the consumption stream by comparing the current consumption with their past standard of living (a habit) which serves as a reference point. Recalling the interpretation of the generator \eqref{intery} from the Introduction it is not surprising to conclude that the generator
\beqo
g(t,Y(t),Z(t))=-u(c(t)-z(t;c_t)),
\eeqo
should model a habit effect depending on the past consumption, see Detemple \& Zapatero \cite{detemple1} and Detemple \& Zapatero \cite{detemple}. However, the reference point does not have to be related to the past consumption. Dybvig \& Rogers \cite{rogers} and K\H{o}szegi \& Rabin \cite{koszegi} advocate to use the investor's past beliefs or past expectations about the pay-off, independent of the past consumption, as the reference point for the valuation. Within such a framework a disappointment effect arises.\\
\indent Let us recall the disappointment model of Loomes and Sugden \cite{loomes}. The one-period utility from the consumption $c$ is valued as
\beqo
Y(0)=\be[u(c)+F(u(c)-H(0))],
\eeqo
where $F$ is an increasing function which measures a disappointment when the realized consumption at time $t=1$ does not match the prior expectation $H(0)=\be[u(c)]$ made at time $t=0$. Extending the model into a continuous dynamic setting we would get
\beqo
Y(t)=\be\Big[\int_t^T\big(u(c(s))+F(s,u(c(s))-H(s))\big)ds|\bfi_t\Big],\quad 0\leq t\leq T.
\eeqo
Alternatively, recalling the interpretation \eqref{intery} we conclude that the disappointment effect from Loomes \& Sugden \cite{loomes} should be modelled by applying the generator
\beq\label{disgen}
g(t,Y(t),Z(t))=-u(c(t))-F(u(c(t))-H(t)).
\eeq
\indent Our goal is to derive a generator $g$ which would give rise to a linear non-monotone preference under a disappointment effect. Following Loomes \& Sugden \cite{loomes}, Dybvig \& Rogers \cite{rogers}, K\H{o}szegi \& Rabin \cite{koszegi} a disappointment effect should be independent of the consumption stream and the reference point $H$ should be fully determined by "the expectation" the investor held in the past. Hence, the past expectations in $H$ must be modelled as an $\bfi_t$-adapted process depending on the path of $Y$. In fact, the reference point $H(t)$ has to depend on the values of $Y$ prior to time $t$. Although a need for a time-delay in the generator should already be obvious at this point, we additionally remark that K\H{o}szegi \& Rabin \cite{koszegi} state clearly that a disappointment model based on the past beliefs must depend on "the lagged expectations". From Rozen \cite{rozen} we learn that a disappointment should lead to non-monotone preferences. In order to model a non-monotone solution to a backward stochastic differential equation we are led again into consideration of a BSDE with a time-delayed generator. At last, we require linearity. As the generator could depend on the whole trajectory of $Y$, the required linearity could take different forms. We could apply the generator of an integral form
\beq\label{bsdeintegral}
g(t,Y_t,Z_t)=-u(c(t))-\beta\big(u(c(t))-\int_0^tY(s)ds\big), \quad \beta>0.
\eeq
In fact, this form of the generator and the corresponding BSDE is investigated in Antonelli et al \cite{antonelli}. We point out that Antonelli et al \cite{antonelli}, in spite of noticing a potential in modelling a disappointment effect, are more interested in explaining an equity puzzle. They do not study properties of the solution to their BSDE and the resulting disappointment effect. We also remark that the BSDE with the time-delayed generator \eqref{bsdeintegral} is investigated from the mathematical point of view in Delong \& Imkeller \cite{delong2}. In this paper we decide to apply the generator of a moving average type
\beq\label{geny}
g(t,Y_t,Z_t)=-u(c(t))-\beta\big(u(c(t))-\frac{1}{t}\int_0^tY(s)ds\big), \quad \beta>0.
\eeq
Under \eqref{geny}, a high average of the past expected utilities requires a high consumption to sustain a high level of the instantaneous utility. We believe that a comparison of the current consumption with the average of the past expectations \eqref{geny} is more reasonable that with the integrated past expectations \eqref{bsdeintegral}. A motivation for choosing a moving average in the generator comes also from the fact that making decisions based on an average seems to have the strongest foundations in economics.\\
\indent Lazrak \& Quenez \cite{lazrak} argue that the generator in a BSDE should depend on the control process $Z$. The authors interpret the generator depending on $Z$ in terms of a risk penalty, and such a dependency with respect to the control process models an aversion against the volatility in the utility process. Motivated by the idea from Loewenstein \& Prelec \cite{loewen2}, we aim at studying a model in which an aversion against volatility causes preferences to be non-monotone. Our goal is to construct a generator which would give rise to a linear non-monotone preference under a volatility aversion. In accordance with \eqref{geny}, we propose to apply the generator
\beq\label{genz}
g(t,Y_t,Z_t)=-u(c(t))-\beta\big(u(c(t))-\frac{1}{t}\int_0^tZ(s)ds\big),\quad \beta>0.
\eeq
Under \eqref{genz} periods of high volatilities penalize the instantaneous utility of the consumption unless the utility of the current consumption is sufficiently large compared to the average of the past volatilities. Referring to the disappointment effect, "a disappointment" may arise when the pay-off is too low compared to the volatility risk which the investor faces. It is clear that the investor requires a sufficient compensation for taking the exposure in the volatility risk. \\
\indent At the end we give one more interpretation of our generators \eqref{geny} and \eqref{genz}. Recalling the interpretation \eqref{intery}, we could say that under the generator \eqref{geny} the price (utility) $Y$ for $\xi$ is assumed to change proportionately to the average of the past observed prices. Under the infinitesimal conditional expected price change $g(t,y,z)=\beta \frac{1}{t}\int_0^ty(s)ds$, high, on average, past prices imply that the investor expects to trade $\xi$ for a high price in the next day. Under the generator \eqref{genz} the price (utility) $Y$ for $\xi$ is assumed to change
proportionately to the average of the past price volatilities. Under the infinitesimal conditional expected price change $g(t,y,z)=\beta \frac{1}{t}\int_0^tz(s)ds$, high, on average, past price volatilities imply that the investor expects to trade $\xi$ for a high price in the next day. These seem to be intuitive feedback relations as far as investors' local beliefs or local pricing rules are concerned.\\
\indent In the sequel, let $\big(\Omega,\mathcal{F},\mathbb{P}, (\mathcal{F}_{t})_{0\leq t\leq T}\big)$ denote a filtered probability space and assume that $\mathbb{F}$ is the natural filtration generated by a Brownian motion $W:=(W(t),0\leq t\leq T)$. We shall work with the following function spaces:
\begin{itemize}
    \item[1.] Let $\mathbb{L}^{p}(\mathbb{R)}$ denote
        the space of $\mathcal{F}_{T}$-measurable random variables $\xi:\Omega\rightarrow\mathbb{R}$
        which fulfill
        $$\mathbb{E}\big[\big|\xi\big|^{p}\big]<\infty.$$
        The space $\mathbb{L}^{p}(\mathbb{R}_+)$ contains $\xi\in\mathbb{L}^{p}(\mathbb{R)}$ which satisfy $\xi\geq 0$ and $\bp(\xi>0)>0$.
    \item [2.] Let $\mathbb{H}^{2}(\mathbb{R)}$ denote
        the space of $\mathbb{F}$-predictable processes $Z:\Omega\times[0,T]\rightarrow\mathbb{R}$
        such that
        $$\mathbb{E}\big[\int_{0}^{T}\big|Z(t)\big|^{2}dt\big]<\infty.$$
    \item [3.] Let $\mathbb{S}^{2}(\mathbb{R)}$ denote
        the space of continuous, $\mathbb{F}$-adapted
        $Y:\Omega\times[0,T]\rightarrow\mathbb{R}$ satisfying
        $$\mathbb{E}\big[\int_0^T\big|Y(t)\big|^{2}\big]<\infty.$$
  \end{itemize}
From the point of view of applications, the random variables $\xi\in\mathbb{L}^p(\mathbb{R}_+)$ are of the greatest importance.\\
\indent In the next sections we show that the BSDEs with the generators of the moving average type, which we have derived in rather a heuristic way, indeed model the properties we require. We omit a consumption stream $c$ and we focus on the terminal pay-off $\xi$. The inclusion of a consumption stream does not change qualitative and quantitative conclusions of this paper.

\section{The BSDE with the moving average generator with respect to the process $Y$}
\indent In this section we investigate the BSDE
\beq\label{bsdey}
Y(t)=\xi-\int_t^T\beta \Big(\frac{1}{s}\int_0^s Y(u)du\Big)ds-\int_t^TZ(s)dW(s),\quad 0\leq t\leq T,
\eeq
with $\beta>0$. The case of $\beta<0$ can be handled analogously.
\begin{prop}
Assume that $\xi\in\mathbb{L}^2(\mathbb{R})$. There exists a unique solution $(Y,Z)\in\mathbb{S}^2(\mathbb{R})\times\mathbb{H}^2(\mathbb{R})$ to the BSDE with the time-delayed generator \eqref{bsdey}. The solution can be represented as
\beq\label{soly}
Y(t)&=&\frac{\be[\xi]}{I_0(2\sqrt{\beta T})}I_0(2\sqrt{\beta t})+\int_0^t \psi(s,t,T)M(s)dW(s),\quad 0\leq t\leq T,\nonumber\\
Z(t)&=&\frac{M(t)}{2I_0(2\sqrt{\beta T})\sqrt{\beta t}K_1(2\sqrt{\beta t})+2K_0(2\sqrt{\beta T})\sqrt{\beta t}I_1(2\sqrt{\beta t})},\quad 0\leq t\leq T,\quad \ \quad
\eeq
where
\beq\label{psiy}
\quad \psi(s,t,T)=\frac{I_0(2\sqrt{\beta t})K_1(2\sqrt{\beta s})+K_0(2\sqrt{\beta t})I_1(2\sqrt{\beta s})}{I_0(2\sqrt{\beta T})K_1(2\sqrt{\beta s})+K_0(2\sqrt{\beta T})I_1(2\sqrt{\beta s})},\quad 0\leq s\leq t\leq T,
\eeq
and the process $M$ is derived form the martingale representation of $\xi$
\beqo
\xi=\be [\xi]+\int_0^TM(s)dW(s).
\eeqo
\end{prop}
\Proof
\noindent \emph{1. The existence and uniqueness of a solution to an integral equation}. We start with investigating a deterministic integral equation corresponding to \eqref{bsdey} of the form
\beq\label{deter}
y(t)=y(0)+\int_0^t\frac{\beta}{s}\int_0^sy(u)duds+f(t),\quad 0\leq t\leq T,
\eeq
with a finite initial condition $y(0)$ and a continuous function $f$ such that $f(0)=0$. We are looking for a continuous solution to \eqref{deter}.\\
\noindent Assume for a moment that $f\in\mathcal{C}^2([0,T])$ with $|f'(0)|<\infty$. By differentiating \eqref{deter} twice, we obtain the second order non-homogeneous differential equation
\beq\label{dif}
ty''(t)+y'(t)-\beta y(t)=h(t),\quad h(t)=tf''(t)+f'(t).
\eeq
In Chapter 2.1 in Polyanin \& Zaitsev \cite{pol} we can find the fundamental solution to the homogeneous equation $ty''(t)+y'(t)-\beta y(t)=0$, which is the function
\beqo
y(t)=\alpha_1I_0(2\sqrt{\beta t})+\alpha_2K_0(2\sqrt{\beta t}),
\eeqo
where $I_0$ and $K_0$ are modified Bessel functions of the first and second kind. We can conclude that the general solution to the non-homogenous equation \eqref{dif} must be of the form
\beq\label{sol1}
y(t)&=&\alpha_1I_0(2\sqrt{\beta t})+\alpha_2K_0(2\sqrt{\beta t})\nonumber\\
&&+2\int_0^th(x)K_0(2\sqrt{\beta x})dxI_0(2\sqrt{\beta t})\nonumber\\
&&-2\int_0^th(x)I_0(2\sqrt{\beta x})dxK_0(2\sqrt{\beta t}).
\eeq
Integrating by parts and applying $h(x)=(xf(x))'$, \eqref{limiti}, \eqref{derivativei}, \eqref{limiki}, \eqref{derivativek}, \eqref{wronskian}, we can obtain the following equivalent representation of \eqref{sol1}
\beq\label{soldeter}
y(t)&=&\alpha_1I_0(2\sqrt{\beta t})+\alpha_2K_0(2\sqrt{\beta t})+f(t)\nonumber\\
&&+2\beta I_0(2\sqrt{\beta t})\int_0^tf(x)K_0(2\sqrt{\beta x})dx\nonumber\\
&&-2\beta K_0(2\sqrt{\beta t})\int_0^tf(x)I_0(2\sqrt{\beta x})dx.
\eeq
As the initial value $y(0)$ must be finite, we have to choose $\alpha_2=0$. To fulfill the initial condition we have to set $\alpha_1=y(0)$, as $I_0(0)=1$. One can show that $y(t)$ given in \eqref{soldeter} is well-defined on $[0,T]$, in particular $t\mapsto y(t)$ is continuous. \\
\noindent We can now check that the solution \eqref{soldeter} to the differential equation \eqref{dif}
satisfies the integral equation \eqref{deter} without the property of differentiability of $f$. Assume that there is another continuous solution $\tilde{y}$ to \eqref{deter}. The function $\hat{y}(t)=y(t)-\tilde{y}(t)$ must satisfy the integral equation
\beqo
\hat{y}(t)=\int_0^t\frac{\beta}{s}\int_0^s\hat{y}(u)duds,\quad 0\leq t\leq T.
\eeqo
We can conclude that $\hat{y}\in\mathcal{C}^{2}((0,T])\cap C([0,T])$ and $\hat{y}$ must fulfill the equation
\beq\label{dif}
t\hat{y}''(t)+\hat{y}'(t)-\beta \hat{y}(t)=0,\quad \hat{y}(0)=0.
\eeq
We obtain that $\hat{y}(t)=0$.\\
\noindent \emph{2. A candidate solution to \eqref{bsdey}}. We substitute $f(t)=\int_0^tZ(s)dW(s)$ in \eqref{soldeter}. A solution to our BSDE \eqref{bsdey} takes the form
\beq\label{sol2}
Y(t)&=&Y(0)I_0(2\sqrt{\beta t})+\int_0^tZ(s)dW(s)\nonumber\\
&&+2\beta I_0(2\sqrt{\beta t})\int_0^t\int_0^xZ(s)dW(s)K_0(2\sqrt{\beta x})dx\nonumber\\
&&-2\beta K_0(2\sqrt{\beta t})\int_0^t\int_0^xZ(s)dW(s)I_0(2\sqrt{\beta x})dx,\quad 0\leq t\leq T.
\eeq
We now rearrange the terms in \eqref{sol2} by changing the order of integration. By the required square integrability of $Z$ and square integrability of $I_0, K_0$ we can apply the Fubini's theorem for stochastic integrals, see Theorem 4.65 in Protter \cite{protter}, which together with \eqref{integrationi}, \eqref{integrationk} yields
\beqo
\lefteqn{Y(t)=Y(0)I_0(2\sqrt{\beta t})+\int_0^tZ(s)dW(s)}\nonumber\\
&&+I_0(2\sqrt{\beta t})\int_0^t\Big(-2\sqrt{\beta t}K_1(2\sqrt{\beta t})+2\sqrt{\beta s}K_1(2\sqrt{\beta s})\Big)Z(s)dW(s)\\
&&-K_0(2\sqrt{\beta t})\int_0^t\Big(2\sqrt{\beta t}I_1(2\sqrt{\beta t})-2\sqrt{\beta s}I_1(2\sqrt{\beta s})\Big)Z(s)dW(s),\quad 0\leq t\leq T.
\eeqo
Applying \eqref{wronskian} we finally obtain
\beq\label{solbeq}
Y(t)&=&Y(0)I_0(2\sqrt{\beta t})\nonumber\\
&&+2I_0(2\sqrt{\beta t})\int_0^t\sqrt{\beta s}K_1(2\sqrt{\beta s})Z(s)dW(s)\nonumber\\
&&+2K_0(2\sqrt{\beta t})\int_0^t\sqrt{\beta s}I_1(2\sqrt{\beta s})Z(s)dW(s),\quad 0\leq t\leq T.
\eeq
We have to choose $Y(0)$ and $Z$ to fulfill the terminal condition
\beqo
\xi&=&Y(0)I_0(2\sqrt{\beta T})\nonumber\\
&&+2I_0(2\sqrt{\beta T})\int_0^T\sqrt{\beta s}K_1(2\sqrt{\beta s})Z(s)dW(s)\nonumber\\
&&+2K_0(2\sqrt{\beta T})\int_0^T\sqrt{\beta s}I_1(2\sqrt{\beta s})Z(s)dW(s).
\eeqo
The martingale representation theorem implies that the only possible choice is
\beqo
Y(0)&=&\frac{\be[\xi]}{I_0(2\sqrt{\beta T})},\nonumber\\
Z(s)&=&\frac{M(s)}{2I_0(2\sqrt{\beta T})\sqrt{\beta s}K_1(2\sqrt{\beta s})+2K_0(2\sqrt{\beta T})\sqrt{\beta s}I_1(2\sqrt{\beta s})},\quad 0\leq s\leq T,
\eeqo
where the process $M$ is derived uniquely from the martingale representation of $\xi$.\\
\noindent \emph{3. Uniqueness of a solution to \eqref{bsdey}}. It follows from points 1 and 2.\\
\noindent \emph{4. Different representations of the candidate solution}. Let $V(t)=\be[\xi|\bfi_t]$. Fix $t\in[0,T]$ and consider $s\mapsto\psi(s,t,T)$ on $[0,t]$. Our solution can be written as
\beqo
Y(t)=\psi(0,t,T)V(0)+\int_0^t\psi(s,t,T)M(s)dW(s),\quad 0\leq t\leq T.
\eeqo
For a fixed $t\in[0,T]$ by It\^{o} product formula we obtain
\beqo
\lefteqn{\psi(t,t,T)\int_0^tM(s)dW(s)}\\
&&=\int_0^t\psi(s,t,T)M(s)dW(s)+\int_0^t\int_0^sM(u)dW(u)\psi'(s,t,T)ds,
\eeqo
which gives us the second representation
\beq\label{representationy}
Y(t)&=&\psi(0,t,T)V(0)+\psi(t,t,T)(V(t)-V(0))-\int_0^t(V(s)-V(0))\psi'(s,t,T)ds\nonumber\\
&=&\psi(t,t,T)V(t)-\int_0^tV(s)\psi'(s,t,T)ds,\quad 0\leq t\leq T.
\eeq
We also derive the third representation
\beq\label{representationy2}
Y(t)&=&\psi(t,t,T)V(t)-\int_0^t(V(s)-V(t))\psi'(s,t,T)ds-\int_0^tV(t)\psi'(s,t,T)ds\nonumber\\
&=&\psi(0,t,T)V(t)-\int_0^t(V(s)-V(t))\psi'(s,t,T)ds,\quad 0\leq t\leq T.
\eeq
\noindent \emph{5. Verification of the candidate solution}. First, we show that $Z$ is square integrable and $\mathbb{F}$-predictable. Consider $v(t)=2\sqrt{\beta t}K_1(2\sqrt{\beta t})$ on $[0,T]$. Notice that by \eqref{limiki} we have $v(0)=1$ and by \eqref{derivativek} we have
\beqo
v'(t)&=&\frac{\sqrt{\beta}}{\sqrt{t}}K_1(2\sqrt{\beta t})-2\sqrt{\beta t}\big(K_0(2\sqrt{\beta t})+\frac{1}{2\sqrt{\beta t}}K_1(2\sqrt{\beta t})\big)\frac{\sqrt{\beta}}{\sqrt{t}}\\
&=&-2\beta K_0(2\sqrt{\beta t})<0.
\eeqo
Hence, $t\mapsto v(t)$ is decreasing and $v(T)>0$. As $I_0(2\sqrt{\beta T})$ and $K_0(2\sqrt{\beta T})$ are strictly positive and $t\mapsto2\sqrt{\beta t}I_1(2\sqrt{\beta t})$ is increasing on $[0,T]$ we conclude
that the denominator in the definition of $Z$ \eqref{soly} is bounded away from zero. As $M$ is square integrable and $\mathbb{F}$-predictable then $Z\in\mathbb{H}^2(\mathbb{R})$ as well.\\
\noindent Next, we show that $Y$ is square integrable, continuous and $\mathbb{F}$-adapted. We investigate the integral $\int_0^t\psi(s,t,T)M(s)dW(s)$ and prove its square integrability. For $t=T$ we obtain $\int_0^TM(s)dW(s)$ and square integrability clearly holds. Fix $t\in(0,T)$. Consider $s\mapsto\psi(s,t,T)$ on $[0,t]$ and notice that $s\rightarrow\psi(s,t,T)$ is continuous. By \eqref{limiti} and \eqref{limiki} we obtain
\beq\label{psi0}
\psi(0,t,T)=\lim_{s\rightarrow 0}\frac{I_0(2\sqrt{\beta t})+K_0(2\sqrt{\beta t})\frac{I_1(2\sqrt{\beta s})}{K_1(2\sqrt{\beta s})}}{I_0(2\sqrt{\beta T})+K_0(2\sqrt{\beta T})\frac{I_1(2\sqrt{\beta s})}{K_1(2\sqrt{\beta s})}}
=\frac{I_0(2\sqrt{\beta t})}{I_0(2\sqrt{\beta T})}>0.
\eeq
The relations \eqref{derivativei}, \eqref{derivativek}, \eqref{wronskian} and some tedious calculations lead to
\beq\label{derivativepsi}
\lefteqn{\frac{d}{ds}\psi(s,t,T)=\psi'(s,t,T)}\nonumber\\
&=&\frac{K_0(2\sqrt{\beta t})I_0(2\sqrt{\beta T})-I_0(2\sqrt{\beta t})K_0(2\sqrt{\beta T})}{\Big(I_0(2\sqrt{\beta T})+K_0(2\sqrt{\beta T})\frac{I_1(2\sqrt{\beta s})}{K_1(2\sqrt{\beta s})}\Big)^2}\Big(\frac{I_1(2\sqrt{\beta s})}{K_1(2\sqrt{\beta s})}\Big)'\nonumber\\
&=&\frac{1}{2s}\frac{K_0(2\sqrt{\beta t})I_0(2\sqrt{\beta T})-I_0(2\sqrt{\beta t})K_0(2\sqrt{\beta T})}{\Big(I_0(2\sqrt{\beta T})K_1(2\sqrt{\beta s})+K_0(2\sqrt{\beta T})I_1(2\sqrt{\beta s})\Big)^2}>0.
\eeq
We remark that the nominator in \eqref{derivativepsi} is strictly positive due to the strict inequality
\beqo
\frac{K_0(2\sqrt{\beta t})}{I_0(2\sqrt{\beta t})}>\frac{K_0(2\sqrt{\beta T})}{I_0(2\sqrt{\beta T})},\quad 0<t<T,
\eeqo
which follows from the monotonicity of $I_0,K_0$. By applying a similar analysis as for the denominator in \eqref{psiy} we can show that the denominator in \eqref{derivativepsi} fulfills
\beqo
0&<&\sqrt{2T}I_0(2\sqrt{\beta T})K_1(2\sqrt{\beta T})\\
&\leq& \sqrt{2s}\Big(I_0(2\sqrt{\beta T})K_1(2\sqrt{\beta s})+K_0(2\sqrt{\beta T})I_1(2\sqrt{\beta s})\Big)\\
&\leq& \frac{1}{\sqrt{2\beta}} I_0(2\sqrt{\beta T})+\sqrt{2T}K_0(2\sqrt{\beta T})I_1(2\sqrt{\beta T}),\quad 0\leq s\leq t,
\eeqo
and the derivative $s\mapsto\psi'(s,t,T)$ is continuous and uniformly bounded away from zero and above on $[0,t]$. We need this result in the sequel. For the proof of this point, we just conclude that for a fixed $t\in(0,T)$ the mapping $s\rightarrow\psi(s,t,T)$ is increasing , continuous and uniformly bounded away from zero and above on $[0,t]$. By boundedness of $\psi$ the stochastic integral $(\int_0^u\psi(s,t,T)M(s)dW(s))_{u\in[0,t]}$ is a square integrable $\mathbb{F}$-adapted martingale and the random variable $\int_0^t\psi(s,t,T)M(s)dW(s)$ is square integrable as well. By using monotonicity of $s\mapsto\psi(s,t,T)$ on $[0,t]$, continuity of $t\mapsto\psi(t,t,T)$ on $[0,T]$ and $\psi(0,0,T)=\frac{1}{I_0(2\sqrt{\beta T})}$ we can prove square integrability of $Y$ as follows
\beqo
\be\Big[\int_0^T|Y(s)|^2ds\Big]&\leq& L\Big(1+\int_0^T\be\Big[\int_0^s|\psi(u,s,T)|^2|M(u)|^2du\Big]ds\Big)\\
&\leq& L\Big(1+\sup_{s\in[0,T]}|\psi(s,s,T)|\be\Big[\int_0^T|M(u)|^2du\Big]\Big)<\infty.
\eeqo
As for any $t\in[0,T]$ the random variable $\int_0^t\psi(s,t,T)M(s)dW(s)$ is $\bfi_t$-measurable the process $Y$ is $\mathbb{F}$-adapted. Continuity of $t\mapsto Y(t)$ follows from the representation \eqref{representationy}.
\cbdu
\indent We now investigate properties of the static and dynamic operator derived from the solution to the BSDE \eqref{bsdey}.
\begin{prop}
Assume that $\xi\in\mathbb{L}^2(\mathbb{R}_+)$. Let the static operator be defined by $\rho(\xi):=Y(0)$, where $Y(0)$ is the initial value of the solution \eqref{soly} to the BSDE with the time-delayed generator \eqref{bsdey} with the terminal condition $\xi$ and a parameter $\beta>0$. The static operator $\rho(\xi)$ satisfies the following properties:
\begin{enumerate}
  \item Linearity: $\rho(\lambda\xi+\theta\eta)=\lambda\rho(\xi)+\theta\rho(\eta)$, $\lambda, \theta \in\mathbb{R}$,
  \item Monotonicity: $\xi\geq \eta\Rightarrow\rho(\xi)\geq \rho(\eta)$,
    \item The bounds hold
    $$\be[\xi] >\rho^{\beta}(\xi)> \be[e^{-\beta T}\xi],$$
  \item The mapping $\beta\mapsto\rho^{\beta}(\xi)$ is continuous, strictly decreasing on $[0,\infty)$ and the limits hold
    $$\lim_{\beta\rightarrow 0}\rho^{\beta}(\xi)=\be[\xi],\quad \lim_{\beta\rightarrow\infty}\rho^{\beta}(\xi)=0.$$
\end{enumerate}
\end{prop}
\Proof
From \eqref{soly} we get $Y(0)=\frac{\be[\xi]}{I_0(2\sqrt{\beta T})}$. The properties are obvious and follow from the properties of $I_0$. The lower bound in point 3 is derived from the relation
\beqo
I_0(2\sqrt{\beta T})=\sum_{k=0}^{\infty}\frac{(\beta T)^k}{k!k!}<\sum_{k=0}^{\infty}\frac{(\beta T)^k}{k!}=e^{\beta T},
\eeqo
which can be found in 9.6.10 in Abramowitz \& Stegun \cite{abramowitz}.
\cbdu
\indent Notice that the lower bound in point 3 is the discounted expected value arising when the classical generator $g(t, Y_t)=\beta Y(t)$ is used. \\
\begin{prop}
Assume that $\xi\in\mathbb{L}^2(\mathbb{R}_+)$. Let the dynamic operator be defined by $\rho_{t,T}(\xi):=Y(t)$, $0\leq t\leq T$, where $Y$ is the solution \eqref{soly} to the BSDE with the time-delayed generator \eqref{bsdey} with the terminal condition $\xi$ and a parameter $\beta>0$. The dynamic operator $(\rho_{t,T}(\xi),0\leq t\leq T)$ satisfies the following properties:
\begin{enumerate}
  \item Linearity: $\rho_{t,T}(\lambda\xi+\theta\eta)=\lambda\rho_{t,T}(\xi)+\theta\rho_{t,T}(\eta)$, $\lambda, \theta\in \mathbb{R}$,
  \item Time-consistency $\rho_{t,s}(\rho_{s,T}(\xi))=\rho_{t,T}(\xi)$, $0\leq t\leq s\leq T$,
  \item If $\xi\geq \eta$ and $\rho_{t,T}(\xi)=\rho_{t,T}(\eta)$ for some $t\in[0,T]$ hold, then $\rho_{s,T}(\xi)=\rho_{s,T}(\eta)$ holds for $0\leq s\leq T$,
  \item The bound holds
  \beqo
  \rho_{t,T}(\xi)\leq\be[\xi|\bfi_t],
  \eeqo
  The bound is strict for $0\leq t<T$,
  \item The mapping $\beta\mapsto \rho_{t,T}^\beta(\xi)$ defined on $[0,\infty)$ is continuous in $\mathbb{L}^2(\mathbb{R})$ and the limits hold
    $$\lim_{\beta\rightarrow 0}\rho^{\beta}_{t,T}(\xi)=\be[\xi|\bfi_t], \quad \lim_{\beta\rightarrow\infty}\rho^{\beta}_{t,T}(\xi)=\mathbf{1}\{t=T\}\xi,$$
  \item Let $\xi, \xi_1, \xi_2, ...$ be a sequence of square integrable random variables. If $\be[|\xi^n-\xi|^2]\rightarrow 0, \ n\rightarrow\infty$, then $\be\big[\int_0^T|\rho_{t,T}(\xi^n)-\rho_{t,T}(\xi)|^2dt\big]\rightarrow 0, \ n\rightarrow \infty$.
\end{enumerate}
\end{prop}
\noindent We write $\rho_t$ unless the terminal time has to be pointed out.\\
\Proof
\noindent 1. Linearity follows easily from the representation \eqref{soly}.\\
\noindent 2. Let $Y(t,T,\xi)$ denote the solution at time $t\in[0,T]$ to the equation \eqref{bsdey} with the terminal condition $\xi$ at time $T$. We have
\beqo
\rho_{t,T}(\xi)&=&Y(t,T, \xi)=\be\Big[\xi-\int_t^T(\frac{1}{u}\int_0^uY(v,T,\xi)dv)du|\bfi_t\Big],\\
\rho_{t,s}(\rho_{s,T}(\xi))&=&Y(t,s,Y(s,T,\xi))\\
&=&\be\Big[Y(s,T,\xi)-\int_t^s(\frac{1}{u}\int_0^uY(v,s,Y(s,T,\xi))dv)du|\bfi_t\Big].
\eeqo
The property of conditional expectations and uniqueness of a solution to our time-delayed BSDE yield that $Y(t,s,Y(s,T,\xi))=Y(t,T,\xi)$.\\
\noindent 3. Recalling \eqref{soly}, the equality $\rho_t(\xi)=\rho_t(\eta)$ for some $t\in[0,T]$ implies that
\beqo
\frac{I_0(2\sqrt{\beta t})}{I_0(2\sqrt{\beta T})}\big(\be[\xi]-\be[\eta]\big)=\int_0^t\psi(s,t,T)\big(M^\eta(s)-M^\xi(s)\big)dW(s),
\eeqo
where $M^\xi, M^\eta$ are the solutions corresponding to the terminal conditions $\xi, \eta$. Taking the expected value we arrive at $\be[\xi-\eta]=0$. As $\xi-\eta\geq 0$ then $\xi=\eta$, and $\rho_{s}(\xi)=\rho_{s}(\eta), 0\leq s\leq T$ holds.\\
\noindent 4. Consider $t\mapsto\psi(t,t,T)$ on $[0,T]$. Applying \eqref{psiy} and \eqref{wronskian} we obtain
\beqo
\psi(t,t,T)=\frac{1}{t\big(I_0(2\sqrt{\beta T})K_1(2\sqrt{\beta t})+K_0(2\sqrt{\beta T})I_1(2\sqrt{\beta t})\big)},\quad 0\leq t\leq T.
\eeqo
The mapping $t\mapsto\psi(t,t,T)$ is continuous with $\psi(0,0,T)=\frac{1}{I_0(2\sqrt{\beta T})}<1$ and $\psi(T,T,T)=1$. Similarly as in \eqref{derivativepsi}, we calculate the derivative
\beqo
\frac{d}{dt}\psi(t,t,T)=\frac{K_0(2\sqrt{\beta t})I_0(2\sqrt{\beta T})-K_0(2\sqrt{\beta T})I_0(2\sqrt{\beta t})}{2t\big(I_0(2\sqrt{\beta T})K_1(2\sqrt{\beta t})+K_0(2\sqrt{\beta T})I_1(2\sqrt{\beta t})\big)^2},
\eeqo
and we conclude that $t\mapsto\psi(t,t,T)$ is strictly increasing and $\frac{1}{I_0(2\sqrt{\beta T})}\leq\psi(t,t,T)\leq1$ holds for $0\leq t\leq T$. The representation \eqref{representationy} and non-negativity of $\xi,\psi'(s,t,T)$, recall \eqref{derivativepsi}, imply now that
\beq\label{upperboundy}
Y(t)\leq \psi(t,t,T)V(t)\leq V(t),\quad 0\leq t\leq T.
\eeq
\noindent 5. Recall that
\beq\label{3ciagi}
\psi(0,t,T)\leq\psi(s,t,T)\leq\psi(t,t,T)\leq1, \quad  0\leq s\leq t.
\eeq
Let $Y^\alpha, Y^\beta$ denote the solutions corresponding to different coefficients in the time-delayed generator. The representation \eqref{soly} gives us the estimate
\beqo
\be[|Y^\beta(t)-Y^\alpha(t)|^2]&\leq& L\Big|\frac{I_0(2\sqrt{\beta t})}{I_0(2\sqrt{\beta T})}-\frac{I_0(2\sqrt{\alpha t})}{I_0(2\sqrt{\alpha T})}\Big|^2\be[|\xi|^2]\\
&&+L\be\Big[\big|\int_0^t\psi^\beta(s,t,T)M(s)dW(s)-\int_0^t\psi^\alpha(s,t,T)M(s)dW(s)\big|^2\Big]\\
&=&L\Big|\frac{I_0(2\sqrt{\beta t})}{I_0(2\sqrt{\beta T})}-\frac{I_0(2\sqrt{\alpha t})}{I_0(2\sqrt{\alpha T})}\Big|^2\be[|\xi|^2]\\
&&+L\be\Big[\int_0^t|\psi^\beta(s,t,T)-\psi^\alpha(s,t,T)|^2|M(s)|^2ds\Big].
\eeqo
The continuity of $\beta\mapsto\rho^\beta_{t}(\xi)$ on $[0,\infty)$ now follows by (uniform in $\beta$) boundedness of $s\mapsto\psi^\beta(s,t,T)$, dominated convergence theorem and continuity of $\beta\mapsto\psi^\beta(s,t,T), \beta\mapsto I_0(2\sqrt{\beta t})$. Notice that $Y^0(t)=\be[\xi]+\int_0^tM(s)dW(s)$ and the limits hold: $\lim_{\beta\rightarrow 0}I_0(2\sqrt{\beta t})=1$, $\lim_{\beta\rightarrow 0}\psi^\beta(0,t,T)=1$, see \eqref{psi0}, and $\lim_{\beta\rightarrow 0}\psi^\beta(s,t,T)=1$, by \eqref{3ciagi}.\\
\noindent The limit $\beta\rightarrow 0$ follows from the continuity. We calculate the limit $\beta\rightarrow\infty$. Choose $t\in[0,T)$. Applying \eqref{limiti} we can show that
\beqo
\lim_{\beta\rightarrow\infty}\psi(0,t,T)= \lim_{\beta\rightarrow\infty}\frac{I_0(2\sqrt{\beta t})}{I_0(2\sqrt{\beta T})}=\lim_{\beta\rightarrow\infty}\frac{e^{2\sqrt{\beta t}}}{\sqrt{2\sqrt{\beta t}}}\frac{\sqrt{2\sqrt{\beta T}}}{e^{2\sqrt{\beta T}}}=0,
\eeqo
and the same limit holds for $\lim_{\beta\rightarrow\infty}\frac{I_1(2\sqrt{\beta t})}{I_0(2\sqrt{\beta T})}$. Applying \eqref{limiki} we can also show that
\beqo
\lim_{\beta\rightarrow\infty}\frac{K_0(2\sqrt{\beta s})}{K_1(2\sqrt{\beta t})}=\lim_{\beta\rightarrow\infty}\frac{e^{-2\sqrt{\beta s}}}{\sqrt{4\sqrt{\beta s}}}\frac{\sqrt{4\sqrt{\beta t}}}{e^{-\sqrt{\beta t}}}=\left\{\begin{array}{ll}
1,\quad s=t\\
0,\quad s=T
\end{array}\right..
\eeqo
Combining the limits we arrive at
\beqo
\lim_{\beta\rightarrow \infty}\psi^\beta(t,t,T)=\frac{\frac{I_0(2\sqrt{\beta t})}{I_0(2\sqrt{\beta T})}+\frac{K_0(2\sqrt{\beta t})}{K_1(2\sqrt{\beta t})}\frac{I_1(2\sqrt{\beta t})}{I_0(2\sqrt{\beta T})}}{1+\frac{K_0(2\sqrt{\beta T})}{K_1(2\sqrt{\beta t})}\frac{I_1(2\sqrt{\beta t})}{I_0(2\sqrt{\beta T})}}=0,
\eeqo
and from \eqref{3ciagi} the convergence of $\lim_{\beta\rightarrow\infty}\psi^\beta(s,t,T)=0$ and $\lim_{\beta\rightarrow\infty}\rho_{t,T}(\xi)$ can be deduced.\\
\noindent 6. Let $(Y^n, Z^n), (Y, Z)$ denote the solutions under the terminal conditions $\xi^n, \xi$. Applying the representation \eqref{soly}, Cauchy-Schwarz inequality, Fubini's theorem, square integrability of $\int_0^t\psi(s,t,T)(Z^n(s)-Z(s))dW(s)$, boundedness of $I_0$ and $\psi$ and the martingale representation of $\xi^n-\xi$
we arrive at
\beqo
\lefteqn{\be\Big[\int_0^T|Y^n(t)-Y(t)|^2dt\Big]}\\
&\leq &L\int_0^T\Big(\be\big[|\xi^n-\xi|^2]+\be\Big[\int_0^t|\psi(s,t,T)|^2|Z^n(s)-Z(s)|^2\Big]ds\Big)dt\\
&\leq &L\int_0^T\Big(\be\big[|\xi^n-\xi|^2]+\be\Big[\int_0^T|Z^n(s)-Z(s)|^2\Big]ds\Big)dt\\
&=& L\int_0^T\Big(\be\big[|\xi^n-\xi|^2]+\be\big[\big|\xi^n-\be[\xi^n]-\xi+\be[\xi]\big|^2\big]\Big)dt\leq L\be\big[|\xi^n-\xi|^2\big],
\eeqo
which proves the convergence.
\cbdu
\indent Some important properties which are fulfilled in the static case do not hold in the dynamic case. Our dynamic operator does not share some key properties which are satisfied by solutions to BSDEs without delays.\\
\noindent \textbf{Example 3.1:} The dynamic operator $\rho_t$ is not monotonic with respect to the terminal condition: $\xi\geq \eta$ may not imply $\rho_t(\xi)\geq \rho_t(\eta), \ 0< t< T$.\\
\noindent Take $\xi=e^{2W(T)-2T}$. Clearly $\xi>0$. From the representation \eqref{representationy} we have that
\beqo
Y(t)=\psi(t,t,T)e^{2W(t)-2t}-\int_0^te^{2W(s)-2s}\psi'(s,t,T)ds,\quad 0\leq t\leq T.
\eeqo
Fix an arbitrary $t\in(0,T)$. As $0<\psi(t,t,T)<1$, the random variable $Y(t)$ can cross zero with a positive probability provided that
\beq\label{monotonicity}
\bp\Big(\int_0^te^{2W(s)-2s}\psi'(s,t,T)ds>e^{2W(t)-2t}\Big)>0.
\eeq
Recall that $s\mapsto\psi'(s,t,T)$ is continuous and bounded away from zero on $[0,t]$, see the discussion after \eqref{derivativepsi}. By applying Theorem 4.1 from Matsumoto \& Yor \cite{yor} for the joint continuous distribution of $(\int_0^{t}e^{2W(s)-2s}ds;W(t))$ on $(0,\infty)\times\mathbb{R}$ we can conclude that
\beqo
\bp\Big(\int_0^te^{2W(s)-2s}ds>\frac{1}{\inf_{0\leq s\leq t} \psi'(s,t,T)}e^{2W(t)-2t}\Big)>0,
\eeqo
which implies \eqref{monotonicity}.
\cbdu
\indent The above example also shows that for $t\in(0,T)$ the monotonicity of the mapping $\beta\mapsto\rho^\beta_t$ fails and the lower bound from point 3 in Proposition 3.2 does not hold for $\rho_t$. As $\rho^\beta_t(\xi)=Y(t)$ can cross zero with positive probability for any $\beta>0$, the operator $\rho^\beta_t$ cannot be dominated a.s. from below by $0$ which is also the limit for $\rho^\beta_t$ arising when $\beta\rightarrow\infty$. Monotonicity of $\beta\mapsto\rho^\beta_t$ for $t\in(0,T)$ fails by recalling that $\rho^0_t(\xi)>0$. As $\rho_t$ is not dominated from below by zero, it is not dominated by the classical discounted expected value operator as well and the lower bound fails.\\
\noindent  \textbf{Example 3.2:} The dynamic operator $\rho_t$ is not conditionally invariant with respect to the terminal condition: $\rho_{s,T}(\xi)=\xi\rho_{s,T}(1), \ t \leq s\leq T, 0< t< T$ for an $\bfi_t$-measurable $\xi$ may not hold.\\
\noindent Choose $t\in (0,T)$ and an $\bfi_t$-measurable, square integrable, non-constant random variable $\xi$. From \eqref{representationy2} we obtain
\beqo
Y(t)=\psi(0,t,T)\xi-\int_0^t(V(s)-\xi)\psi'(s,t,T)ds=\rho_{t,T}(1)\xi-\int_0^t(V(s)-\xi)\psi'(s,t,T)ds,
\eeqo
and the last integral does not vanish. Choose $\xi=\mathbf{1}\{W(t)>0\}$ and we get $V(s)=\Phi(\frac{W(s)}{\sqrt{t-s}}), \ 0\leq s \leq t$, where $\Phi$ is a distribution function of a standard normal distribution. The integral in this case is strictly positive or negative depending on the final value of $\xi$.
\cbdu

\section{The BSDE with the moving average generator with respect to the process $Z$}
\indent In this section we deal with the time-delayed BSDE
\beq\label{bsdez}
Y(t)=\xi-\int_t^T\beta \Big(\frac{1}{s}\int_0^s Z(u)du\Big)ds-\int_t^TZ(s)dW(s),\quad 0\leq t\leq T,
\eeq
with $\beta>0$. The case of $\beta<0$ can be handled analogously.
\begin{prop}
Assume that $\xi\in\mathbb{L}^{2+\epsilon}(\mathbb{R})$ for some $\epsilon>0$. We introduce the equivalent probability measure $\bqq\sim\bp$
\beq\label{measurez}
\frac{d\tilde{\bq}}{d\bp}\big|\bfi_T=N^\beta(T)=e^{-\int_0^T\beta\ln(\frac{T}{s})dW(s)-\frac{1}{2}\int_0^T\big(\beta\ln(\frac{T}{s})\big)^2ds}.
\eeq
Consider the process $Y$ defined as
\beq\label{solz}
Y(t)&=&\be^{\bqq}[\xi|\bfi_t]-\beta\ln(\frac{T}{t})\int_0^t Z(s)ds,\quad 0\leq t\leq T,
\eeq
and the process $Z$ derived from the martingale representation
\beqo
\xi=\be^{\bqq}[\xi]+\int_0^TZ(s)d\tilde{W}(s),
\eeqo
where $\tilde{W}$ is a $\bqq$-Brownian motion. Under the assumption that $\sup_{0\leq t\leq T}|Z(t)|<\infty$, the pair of processes $(Y,Z)\in\mathbb{S}^2(\mathbb{R})\times\mathbb{H}^2(\mathbb{R})$ is a unique solution to the BSDE with the time-delayed generator \eqref{bsdez}.
\end{prop}
\noindent \textbf{Remark:} The required a.s. finiteness of $Z$ ensures that the integral $\int_t^T\frac{1}{s}\int_0^s Z(u)duds$ appearing in \eqref{bsdez} exists \emph{a.s.}.\\
\Proof
\noindent \emph{1. A candidate solution to \eqref{bsdez}}. We change the order of integration and we obtain
\beq\label{eqqq1}
Y(t)&=&\xi-\beta\ln(\frac{T}{t})\int_0^t Z(s)ds\nonumber\\
&&-\int_t^T\beta\ln(\frac{T}{s})Z(s)ds-\int_t^TZ(s)dW(s), \quad 0\leq t\leq T.
\eeq
We introduce the equivalent probability measure $\tilde{\bq}$ and we rewrite \eqref{eqqq1} as
\beq\label{bsdezz}
Y(t)=\xi-\beta\ln(\frac{T}{t})\int_0^t Z(s)ds-\int_t^TZ(s)d\tilde{W}(s),\quad 0\leq t\leq T,
\eeq
where $\tilde{W}$ is a Brownian motion under $\bqq$. The form of a solution follows by taking the conditional expected value.\\
\noindent \emph{2. The uniqueness of a solution to \eqref{bsdez}}. Assume there are two solutions $(Y^1,Z^1),(Y^2,Z^2)$. By subtracting the relation \eqref{bsdezz} for these two solutions we arrive at
\beqo
Y^1(0)-Y^2(0)=\int_0^T(Z^1(s)-Z^2(s))d\tilde{W}(s).
\eeqo
Taking the expected value we end up with $Y^1(0)=Y^2(0)$. Taking the square and the expectation we conclude that the processes $Z^1, Z^2$ must coincide \emph{a.e.a.s.}.\\
\noindent 3. \emph{Verification of the candidate solution}. We show that $Z$ is square integrable and $\mathbb{F}$-predictable. As $\be[|\xi|^{2+\epsilon}]<\infty$ holds, we have that
\beq\label{integrabilitypower}
\be^\bqq[|\xi|^{2+\frac{\epsilon}{2}}]\leq\big(\be[|N^\beta(T)|^{\frac{2+\epsilon}{\frac{\epsilon}{2}}}]\big)^{\frac{\frac{\epsilon}{2}}{2+\epsilon}}\big(\be[|\xi|^{2+\epsilon}]\big)^{\frac{2+\frac{\epsilon}{2}}{2+\epsilon}}<\infty,
\eeq
and we can conclude from Theorem 5.1 in El Karoui et al \cite{K} that
\beq\label{qintegrabilityz}
\be^\bqq\Big[\Big(\int_0^T|Z(s)|^2ds\Big)^{1+\frac{\epsilon}{4}}\Big]<\infty.
\eeq
Square integrability of $Z$ under $\bp$ can be now established from the estimate
\beq\label{squarezz}
\lefteqn{\be\Big[\int_0^T|Z(s)|^2ds\Big]}\nonumber\\
&\leq& \Big(\be^\bqq\big[|N^\beta(T)|^{-\frac{1+\frac{\epsilon}{4}}{\frac{\epsilon}{4}}}\big]\Big)^{\frac{\frac{\epsilon}{4}}{1+\frac{\epsilon}{4}}}\Big(\be^\bqq\Big[\Big(\int_0^T|Z(s)|^2ds\Big)^{1+\frac{\epsilon}{4}}\Big]\Big)^{\frac{1}{1+\frac{\epsilon}{4}}}<\infty.
\eeq
Clearly, the process $Z$ derived from the martingale representation is $\mathbb{F}$-predictable.\\
\noindent We show that $Y$ is square integrable, continuous and $\mathbb{F}$-adapted. By applying \eqref{solz}, Cauchy-Schwarz inequality, integrability of $\ln^{2}(\frac{T}{s})$, Fubini's theorem, the property of conditional expectations and convexity of the power function we can derive
\beq\label{estimateyy}
\lefteqn{\be^\bqq\Big[\Big(\int_0^T|Y(s)|^{2}ds\Big)^{1+\frac{\epsilon}{4}}\Big]}\nonumber\\
&\leq&L\be^\bqq\Big[\Big(\int_0^T\Big|\be^\bqq[|\xi|^2|\bfi_s]+\ln^2(\frac{T}{s})\int_0^T|Z(u)|^2du\Big|ds\Big)^{1+\frac{\epsilon}{4}}\Big]\nonumber\\
&\leq& L\Big(\be^\bqq\big[|\xi|^{2+\frac{\epsilon}{2}}\big]+\be^\bqq\Big[\Big(\int_0^T|Z(u)|^2du\Big)^{1+\frac{\epsilon}{4}}\Big]\Big)<\infty.
\eeq
Square integrability of $Y$ under $\bp$ can be proved as in \eqref{squarezz}. Clearly, the process $Y$ is $\mathbb{F}$-adapted. Continuity of $t\mapsto Y(t)$ follows from the representation \eqref{solz}. In particular, the requirement $\sup_{0\leq t\leq T}|Z(t)|<\infty$ yields that $\lim_{t\rightarrow 0}\ln(\frac{t}{T})\int_0^tZ(s)ds=0$.
\cbdu
\indent In the sequel we assume that $Z(t)\geq 0, \ 0\leq t\leq T$. Non-negativity of $Z$ allows us to interpret the control process $Z$ as the intensity of variability of $Y$. Before we move further, let us comment on our requirements that $Z(t)\geq 0$ and $\sup_{0\leq t\leq T}|Z(t)|<\infty$. For a large class of pay-offs $\xi$ (especially in financial applications) the process $Z$ derived from the martingale representation $\xi$ is indeed non-negative and a.s. finite on $[0,T]$. This holds for pay-offs of the form $\xi=F(W(T))$ or $\xi=F(\int_0^Th(s)dW(s))$, with a non-negative left continuous and with right limits deterministic function $h$ and an increasing mapping $x\mapsto F(x)$ (together with some additional requirements on $F$).

\begin{prop}
Assume that $\xi\in\mathbb{L}^{2+\epsilon}(\mathbb{R}_+)$ for some $\epsilon>0$. Let the static operator be defined by $\rho(\xi):=Y(0)$, where $Y(0)$ is the initial value of the solution \eqref{solz} to the BSDE with the time-delayed generator \eqref{bsdez} with the terminal condition $\xi$ and a parameter $\beta>0$ under the assumptions that $\sup_{0\leq t\leq T}|Z(t)|<\infty$ and $Z(t)\geq 0, \ 0\leq t\leq T$. The static operator $\rho(\xi):=Y(0)$ satisfies the following properties:
\begin{enumerate}
  \item Linearity: $\rho(\lambda\xi+\theta\eta)=\lambda\rho(\xi)+\theta\rho(\eta)$, $\lambda, \theta\in\mathbb{R}$,
  \item Monotonicity: $\xi\geq \eta\Rightarrow\rho(\xi)\geq \rho(\eta)$,
    \item The bounds hold
    $$\be[\xi]\geq\rho^{\beta}(\xi)\geq -||-\xi||^\infty,$$
    where $||.||^\infty$ denotes the essential supremum of a random variable,
    \item The mapping $\beta\mapsto\rho^{\beta}(\xi)$ is continuous, non-increasing on $[0,\infty)$ and the limits hold
    $$\lim_{\beta\rightarrow 0}\rho^{\beta}(\xi)=\be[\xi], \quad \lim_{\beta\rightarrow\infty}\rho^{\beta}(\xi)=-||-\xi||^\infty.$$
\end{enumerate}
The bounds in point 3 are strict and
the mapping $\beta\mapsto\rho(\beta)$ is strictly decreasing unless $Z(t)=0$ $a.e., a.s.$
\end{prop}
\Proof
Notice that $Y(0)=\be^{\bqq}[\xi]=\be[N^\beta(T)\xi]$ with the measure $\bqq$ defined by $N^\beta$ in \eqref{measurez}. \\
\noindent 1-2. The statements are obvious.\\
\noindent 3-4. We start with proving continuity. Fix $\alpha\geq 0$. The sequence $(N^\beta(T)\xi)_{\beta\in[\alpha-r \vee 0,\alpha+r]}$, for some $r>0$, is uniformly integrable under $\bp$ as
\beqo\
\be\big[|N^\beta(T)\xi|^{1+\frac{\epsilon}{2}}\big]\leq \Big(\be\big[|N^\beta(T)|^{2+\epsilon}\big]\Big)^{\frac{1}{2}}\Big(\be\big[|\xi|^{2+\epsilon}\big]\Big)^{\frac{1}{2}}\leq L,\quad \beta\in[\alpha-r \vee 0,\alpha+r], \epsilon>0.
\eeqo
The limit $\lim_{\beta\rightarrow \alpha}Y^\beta(0)=\lim_{\beta\rightarrow \alpha}\be[N^\beta(T)\xi]=\be[N^{\alpha}(T)\xi]=Y^{\alpha}(0)$ is established by Lebesgue's theorem and a.s. continuity of $\beta\mapsto N^\beta(T)$. \\
\noindent We now move to monotonicity. From \eqref{bsdezz} we have
\beq\label{monotonocity}
Y(0)=\xi-\int_0^T\beta\ln(\frac{T}{s})Z(s)ds-\int_0^TZ(s)dW(s).
\eeq
We denote by $\bqq^\alpha, \bqq^\beta$ the equivalent probability measures \eqref{measurez} induced by $N^\alpha$ and $N^\beta$ with the parameters $\alpha$ and $\beta$. Let $(Y^\alpha, Z^\alpha), (Y^\beta, Z^\beta)$ denote the corresponding solutions. We arrive at
\beqo
\lefteqn{Y^\alpha(0)-Y^\beta(0)}\\
&=&\int_0^T\beta\ln(\frac{T}{s})Z^\beta(s)ds-\int_0^T\alpha\ln(\frac{T}{s})Z^\alpha(s)ds-\int_0^T(Z^\alpha(s)-Z^\beta(s))dW(s)\\
&=& \int_0^T(\beta-\alpha)\ln(\frac{T}{s})Z^\beta(s)ds-\int_0^T(Z^\alpha(s)-Z^\beta(s))d\tilde{W}^\alpha(s),
\eeqo
where $\tilde{W}^\alpha$ is a Brownian motion under $\bqq^\alpha$. As \eqref{qintegrabilityz} holds we immediately conclude that $Z^\alpha$ is square integrable under $\bqq^\alpha$. Square integrability of $Z^\beta$ under $\bqq^\alpha$ follows from
\beq\label{cauchy}
\lefteqn{\be^{\bqq^\alpha}\Big[\int_0^T|Z^\beta(s)|^2ds\Big]=\be^{\bqq^\beta}\Big[\frac{N^\alpha(T)}{N^\beta(T)}\int_0^T|Z^\beta(s)|^2ds\Big]}\nonumber\\
&\leq &\Big(\be^{\bqq^\beta}\Big[\Big|\frac{N^\alpha(T)}{N^\beta(T)}\Big|^{\frac{1+\frac{\epsilon}{4}}{\frac{\epsilon}{4}}}\Big]\Big)^{\frac{\frac{\epsilon}{4}}{1+\frac{4}{\epsilon}}}\Big(\be^{\bqq^\beta}\Big[\Big(\int_0^T|Z^\beta(s)|^2ds\Big)^{1+\frac{\epsilon}{4}}\Big]\Big)^{\frac{1}{1+\frac{\epsilon}{4}}}<\infty.
\eeq
Taking the expected value under $\bqq^\alpha$ in \eqref{monotonocity} we obtain that
\beqo
Y^\alpha(0)-Y^\beta(0)=(\beta-\alpha)\be^{\bqq^\alpha}\Big[\int_0^T\ln(\frac{T}{s})Z^\beta(s)ds\Big],
\eeqo
as the expected value of the stochastic integral vanishes. By Cauchy-Schwarz inequality we get that $\be^{\bqq^\alpha}[\int_0^T\ln(\frac{T}{s})Z^\beta(s)ds]<\infty$. We conclude that $\alpha>\beta$ implies $Y^\alpha(0)\leq Y^\beta(0)$ under the requirement that $Z^\alpha, Z^\beta\geq 0$. The inequality is strict unless $Z(t)=0$ a.e., a.s.. \\
\noindent We finally calculate the limits. The limit $Y^0(0)=\lim_{\beta\rightarrow 0}Y^\beta(0)=\be[\xi]$ is obvious. As the mapping $\beta\mapsto Y^\beta(0)$ is non-increasing and uniformly bounded below by zero, due to the assumption $\xi\geq0$, we can deduce the existence of the limit $Y^\infty(0)=\lim_{\beta\rightarrow\infty}Y^\infty(0)=\inf_{\beta>0}\be^{\bqq^\beta}[\xi]=-||-\xi||^\infty$.
\cbdu
\indent We remark that the lower bound in point 3 in Proposition 4.2 is weaker than the lower bound in point 3 in Proposition 3.2. In contrast to point 3 in Proposition 3.2 we cannot compare our new operator $\rho(\xi)=\be^\bqq[\xi]$ with the expected value under the equivalent probability measure $\bq^{\star}$
\beqo
\frac{d\bq^{\star}}{d\bp}\big|\bfi_T=e^{-\beta W(T)-\frac{1}{2}\beta^2T},
\eeqo
which arises if the classical generator $g(t,Z_t)=\beta Z(t)$ is used.\\
\noindent \textbf{Example 4.1:} We cannot establish a relation between $\rho^\star(\xi)=\be^{\bq^{\star}}[\xi]$ and $\rho(\xi)=\be^\bqq[\xi]$. Let us consider
$\xi=e^{\int_0^Th(s)dW(s)}$ with a non-negative, left-continuous and with right limits deterministic function $h$. We remark that in this case the control process $Z$ arising from \eqref{bsdez} satisfies $\sup_{0\leq t\leq T}|Z(t)|<\infty$ and $Z(t)\geq 0, \ 0\leq t\leq T$. We can find that
\beqo
\rho(\xi)&=&\be^{\bqq}\big[e^{\int_0^Th(s)dW(s)}\big]=\be^{\bqq}\big[e^{\int_0^Th(s)d\tilde{W}(s)-\int_0^T\beta\ln(\frac{T}{s})h(s)ds}\big]\\
&&=e^{\frac{1}{2}\int_0^Th^2(s)ds}e^{-\int_0^T\beta\ln(\frac{T}{s})h(s)ds},\\
\rho^\star(\xi)&=&\be^{\bq^{\star}}\big[e^{\int_0^Th(s)dW(s)}\big]=\be^{\bq^{BS}}\big[e^{\int_0^Th(s)dW^{\bq^{\star}}(s)-\int_0^T\beta h(s)ds}\big]\\
&&=e^{\frac{1}{2}\int_0^Th^2(s)ds}e^{-\int_0^T\beta h(s)ds}.
\eeqo
Take $h(s)=1$ and we obtain $\rho(\xi)=\rho^\star(\xi)$, take $h(s)=\mathbf{1}\{s>\frac{T}{e}\}$) and we obtain $\rho(\xi)>\rho^\star(\xi)$, finally take $h(s)=\mathbf{1}\{s\leq\frac{T}{e}\}$ and we obtain $\rho(\xi)<\rho^\star(\xi)$. \cbdu
\indent Let us move to the dynamic case.
\begin{prop}
Assume that $\xi\in\mathbb{L}^{2+\epsilon}(\mathbb{R}_+)$ for some $\epsilon>0$. Let the dynamic operator be defined by $\rho_{t,T}(\xi):=Y(t)$, $0\leq t\leq T$, where $Y(t)$ is the solution \eqref{solz} to the BSDE with the time-delayed generator \eqref{bsdez} with the terminal condition $\xi$ and a parameter $\beta>0$ under the assumption that $\sup_{0\leq t\leq T}|Z(t)|<\infty$ and $Z(t)\geq 0, \ 0\leq t\leq T$. The dynamic operator $(\rho_{t,T}(\xi), 0\leq t\leq T)$ satisfies the following properties:
\begin{enumerate}
  \item Linearity: $\rho_{t,T}(\lambda\xi+\theta\eta)=\lambda\rho_{t,T}(\xi)+\theta\rho_{t,T}(\eta)$, $\lambda, \theta\in\mathbb{R}$,
  \item Time-consistency $\rho_{t,s}(\rho_{s,T}(\xi))=\rho_{t,T}(\xi)$, $0\leq t\leq s\leq T$,
  \item if $\xi\geq \eta$ and $\rho_{t,T}(\xi)=\rho_{t,T}(\eta)$ for some $t\in[0,T]$ hold, then $\rho_{s,T}(\xi)=\rho_{s,T}(\eta)$ holds for $0 \leq s\leq T$,
  \item The bound holds
  \beqo
  \rho_{t,T}(\xi)\leq \be[\xi|\bfi_t].
  \eeqo
  The bound is strict for $0\leq t<T$ unless $Z(t)=0, a.e., a.s.$,
  \item The mapping $\beta\mapsto\rho_{t,T}(\xi)$ defined on $[0,\infty)$ is continuous in $\mathbb{L}^2(\mathbb{R})$ and the limits hold
    \beqo
    \lim_{\beta\rightarrow 0}\rho^{\beta}_{t,T}(\xi)=\be[\xi|\bfi_t],\\
    \underline{M}^\beta(t)\leq\rho^\beta_{t,T}(\xi)\leq \overline{M}^\beta(t),
    \eeqo
    with
    \beqo
    \lim_{\beta\rightarrow\infty}\overline{M}^{\beta}(t)&=&-||-\xi||^{\infty}_{t}, \text{weakly (and strongly) in } \mathbb{L}^2(\Omega, \bfi_t, \bp;\mathbb{R}),\quad 0\leq t\leq T,\\
    \lim_{\beta\rightarrow\infty}\underline{M}^{\beta}(t)&=&-||-\xi||^{\infty}_{0}, \text{weakly in } \mathbb{L}^2(\Omega, \bfi_t, \bp;\mathbb{R}),\quad 0\leq t\leq T,
    \eeqo
    where $||.||^\infty_{t}$ is the essential supremum of a random variable under the conditional probability. The limits $\beta\rightarrow\infty$ hold provided that $\be\big[\sup_{t\in[0,T]}\big|||-\xi||_t\big|^2\big]<\infty$,
  \item Let $\xi, \xi_1, \xi_2, ...$ be a sequence of random variables having finite $(2+\epsilon)$-th moment, for some $\epsilon>0$. If $\be[|\xi^n-\xi|^{2+\epsilon}]\rightarrow 0, \ n\rightarrow\infty$, then $\be\big[\int_0^T|\rho_{t,T}(\xi^n)-\rho_{t,T}(\xi)|^2dt\big]\rightarrow 0, \ n\rightarrow \infty$.
\end{enumerate}
\end{prop}
\noindent \textbf{Remark:} If the process $t\mapsto||-\xi||^{\infty}_{t}$ is constant on $[0,T)$, then the weak convergence for $\lim_{\beta\rightarrow\infty}\rho^\beta_{t,T}(\xi)$ is proved. In general, strong convergence cannot be established, see Peng \cite{penglimit}.\\
\Proof
\noindent 1. The statement follows easily from the representation \eqref{solz}.\\
\noindent 2. Time-consistency could be proved as in point 2 in Proposition 3.2.\\
\noindent 3. The case of $t=0$ is trivial. Fix $t\in(0,T]$. Notice that $\rho_{t}(\xi)=\rho_{t}(\eta)$ implies that
\beqo
\be^{\bqq}[\xi|\bfi_t]-\beta\ln(\frac{T}{t})\int_0^t Z^\xi(s)ds=\be^{\bqq}[\eta|\bfi_t]-\beta\ln(\frac{T}{t})\int_0^t Z^\eta(s)ds,
\eeqo
with $Z^\xi, Z^\eta$ denoting the corresponding solutions. From the martingale representations of $\xi$ and $\eta$ we have that
\beqo
\lefteqn{\be^\bqq[\xi]+\int_0^tZ^\xi(s)d\tilde{W}(s)-\beta\ln(\frac{T}{t})\int_0^t Z^\xi(s)ds}\\
&&=\be^\bqq[\eta]+\int_0^tZ^\eta(s)d\tilde{W}(s)-\beta\ln(\frac{T}{t})\int_0^t Z^\eta(s)ds.
\eeqo
Rewriting the above equation we obtain
\beq\label{localmart}
\be^\bqq[\xi-\eta]=\int_0^t(Z^\eta(s)-Z^\xi(s))dW^{\bq'}(s),
\eeq
where $W^{\bq'}$ is a Brownian motion under the equivalent probability measure
\beqo
\frac{d\mathbb{Q}'}{d\bqq}\Big|\bfi_T=e^{\beta\ln(\frac{t}{T})\tilde{W}(T)-\frac{1}{2}(\beta\ln(\frac{t}{T}))^2T}.
\eeqo
As in \eqref{cauchy} we can prove that $(\int_0^u(Z^\eta(s)-Z^\xi(s))dW^{\bq'}(s))_{u\in[0,T]}$ is a square integrable martingale under $\bq'$ and the result follows by taking the expected value in \eqref{localmart}.\\
\noindent
\noindent 4. Notice that the solution \eqref{bsdezz} could be rewritten as
\beq\label{twofactors}
Y^\beta(t)&=&X^\beta(t)-\beta\ln(\frac{T}{t})\int_0^tZ^\beta(s)ds,\quad 0\leq t\leq T\nonumber\\
X^\beta(t)&=&\xi-\int_t^T\beta\ln(\frac{T}{s})Z^\beta(s)ds-\int_t^TZ^\beta(s)dW(s),\quad 0\leq t\leq T.
\eeq
We have that $X^\beta(t)=\be^{\bqq^\beta}[\xi|\bfi_t]$. As in the proof of points 3-4 in Proposition 4.2 we can show that for each $t\in[0,T]$ the mapping $\beta\mapsto X^\beta(t)$ is non-increasing and $\lim_{\beta\rightarrow 0}X^\beta(t)=\be[\xi|\bfi_t]$ holds. By non-negativity of $Z$, we obtain the upper bound for $\rho_t$ immediately
\beqo
\rho^{\beta}_{t}(\xi)=X^\beta(t)-\beta\ln(\frac{T}{t})\int_0^tZ^\beta(s)ds\leq \be[\xi|\bfi_t],\quad 0\leq t\leq T.
\eeqo
\noindent 5. Fix $\alpha\geq 0$. We would take $\beta\rightarrow\alpha$. Denote by $(Y^\alpha, Y^\beta)$ the corresponding solutions. Continuity for $t=0$ is proved in Proposition 4.2. Choose $t\in(0,T)$. The relation \eqref{twofactors} and Cauchy-Schwarz inequality yield that
\beq\label{continuityy}
\be\big[|Y^\alpha(t)-Y^\beta(t)|^2\big]&\leq& L\Big(\be[|X^\alpha(t)-X^\beta(t)|^2]\nonumber\\
&&+\beta^2\ln^2(\frac{T}{t})\be\big[\int_0^T|Z^\alpha(s)-Z^\beta(s)|^2ds\big]\nonumber\\
&&+(\alpha-\beta)^2\ln^2(\frac{T}{t})\be\big[\int_0^T|Z^\alpha(s)|^2ds\big]\Big).
\eeq
The last term in \eqref{continuityy} converges to zero when $\beta\rightarrow\alpha$. We have to prove the convergence of the first two terms. By It\^{o} formula we obtain
\beqo
\lefteqn{d\big((X^\alpha(s)-X^\beta(s))^2\big)=2(X^\alpha(s)-X^\beta(s))\big(\alpha\ln(\frac{T}{s})Z^\alpha(s)-\beta\ln(\frac{T}{s})Z^\beta(s)\big)ds}\\
&&+2(X^\alpha(t)-X^\beta(s))\big(Z^\alpha(s)dW(s)-Z^\beta(s)dW(s)\big)+(Z^\alpha(s)-Z^\beta(s))^2ds,
\eeqo
rearranging the terms and changing the measure to $\tilde{\bq}^\beta$ with the martingale $N^\beta$ we arrive at
\beq\label{prioriestimates}
\lefteqn{\int_t^T|Z^\alpha(s)-Z^\beta(s)|^2ds+|X^\alpha(t)-X^\beta(t)|^2}\nonumber\\
&=&-2\int_t^T(X^\alpha(s)-X^\beta(s))(\alpha-\beta)\ln(\frac{T}{s})Z^\alpha(s)ds\nonumber\\
&&-2\int_t^T(X^\alpha(t)-X^\beta(s))(Z^\alpha(s)-Z^\beta(s))d\tilde{W}^\beta(s).
\eeq
By taking the conditional expected value we conclude that
\beqo
|X^\alpha(t)-X^\beta(t)|^2\leq \be^{\bq^\beta}\Big[2\int_0^T|X^\alpha(s)-X^\beta(s)||\alpha-\beta|\ln(\frac{T}{s})Z^\beta(s)ds|\bfi_t\Big],
\eeqo
and by Doob's martingale inequality, Cauchy-Schwarz inequality and integrability of $\ln^2(\frac{T}{s})$ we derive for a sufficiently small $p>1$
\beq\label{intermediateestimate}
\lefteqn{\be^{\bq^{\beta}}\big[\sup_{t\in[0,T]}|X^\alpha(t)-X^\beta(t)|^{2p}\big]}\nonumber\\
&\leq& L\be^{\bq^{\beta}}\Big[\Big|\int_0^T|X^\alpha(s)-X^\beta(s)||\alpha-\beta|\ln(\frac{T}{s})Z^\beta(s)ds\Big|^p\Big]\nonumber\\
&\leq&L\be^{\bq^{\beta}}\Big[\sup_{s\in[0,T]}|X^\alpha(s)-X^\beta(s)|^p\Big|\int_0^T|\alpha-\beta|\ln(\frac{T}{s})Z^\beta(s)ds\Big|^p\Big]\nonumber\\
&\leq&LC^2\be^{\bq^{\beta}}\Big[\sup_{s\in[0,T]}|X^\alpha(s)-X^\beta(s)|^{2p}\Big]\nonumber\\
&&+L\frac{1}{C^2}|\alpha-\beta|^{2p}\be^{\bq^{\beta}}\Big[\Big|\int_0^T\ln(\frac{T}{s})Z^\alpha(s)ds\Big|^{2p}\Big]\nonumber\\
&\leq&LC^2\be^{\bq^{\beta}}\Big[\sup_{s\in[0,T]}|X^\alpha(s)-X^\beta(s)|^{2p}\Big]\nonumber\\
&&+L\frac{1}{C^2}|\alpha-\beta|^{2p}\be^{\bq^{\beta}}\Big[\Big|\int_0^T|Z^\alpha(s)|^2ds\Big|^{p}\Big].
\eeq
Choosing $C$ sufficiently small we obtain the estimate
\beq\label{prior1x}
\be^{\bq^{\beta}}\big[\sup_{t\in[0,T]}|X^\alpha(t)-X^\beta(t)|^{2p}\big]\leq L|\alpha-\beta|^{2p}\be^{\bq^{\beta}}\Big[\Big|\int_0^T|Z^\alpha(s)|^2ds\Big|^{p}\Big].
\eeq
By recalling \eqref{prioriestimates}, applying the convexity property of the power function and Burkholder inequality give us for a sufficiently small $p>1$
\beqo
\lefteqn{\be^{\bq^\beta}\Big[\Big(\int_0^T|Z^\alpha(s)-Z^\beta(s)|^2ds\Big)^p\Big]}\\
&\leq& L\be^{\bq^{\beta}}\Big[\Big(\int_0^T|X^\alpha(s)-X^\beta(s)|^2|Z^\alpha(s)-Z^\beta(s)|^2ds\Big)^{p/2}\Big]\\
&&+L\be^{\bq^{\beta}}\Big[\Big|\int_0^T|X^\alpha(s)-X^\beta(s)||\alpha-\beta|\ln(\frac{T}{s})Z^\alpha(s)ds\Big|^p\Big]\\
&\leq&L\be^{\bq^{\beta}}\Big[\sup_{s\in[0,T]}|X^\alpha(s)-X^\beta(s)|^p\Big(\int_0^TZ^\alpha(s)-Z^\beta(s)|^2ds\Big)^{p/2}\Big]\\
&&+L\be^{\bq^{\beta}}\Big[\sup_{s\in[0,T]}|X^\alpha(s)-X^\beta(s)|^p\Big|\int_0^T|\alpha-\beta|\ln(\frac{T}{s})Z^\alpha(s)ds\Big|^p\Big].
\eeqo
Following \eqref{intermediateestimate}, together with \eqref{prior1x}, we obtain
\beq\label{prior2z}
\be^{\bq^{\beta}}\Big[\Big(\int_0^T|Z^\alpha(t)-Z^\beta(t)|^{2}\Big)^p\big]\leq L|\alpha-\beta|^{2p}\be^{\bq^{\beta}}\Big[\Big|\int_0^T|Z^\alpha(s)|^2ds\Big|^{p}\Big].
\eeq
The estimates \eqref{prior1x} and \eqref{prior2z} allow us to prove the convergence of the two terms on the right hand side of \eqref{continuityy}. Choose $p>1$ and $a>1$ sufficiently small. We could derive by applying H\"{o}lder's inequalities
\beqo
\be\big[|X^\alpha(t)-X^\beta(t)|^2\big]&\leq& \Big(\be^{\bq^\beta}\big[|X^\alpha(t)-X^\beta(t)|^{2p}\big]\Big)^{\frac{1}{p}}\Big(\be^{\bq^\beta}[\big|N^\beta(T)\big|^{-q}]\Big)^{\frac{1}{q}}\\
&\leq&L|\alpha-\beta|^{2}\Big(\be^{\bq^\beta}\Big[\big|\int_0^T|Z^\alpha(s)|^2ds\big|^{p}\Big]\Big)^{\frac{1}{p}}\Big(\be^{\bq^\beta}[\big|N^\beta(T)\big|^{-q}]\Big)^{\frac{1}{q}}\\
&\leq&L|\alpha-\beta|^{2}\Big(\be\Big[\Big|\int_0^T\big|Z^\alpha(s)|^2ds\big|^{ap}\Big]\Big)^{\frac{1}{ap}}\Big(\be[\big|N^\beta(T)\big|^{b}]\Big)^{\frac{1}{bp}}\\
&&\cdot\Big(\be[\big|N^\beta(T)\big|^{-2q}]\Big)^{\frac{1}{2q}}\Big(\be[\big|N^\beta(T)\big|^{2}]\Big)^{\frac{1}{2q}}\\
&=&G(\beta)L|\alpha-\beta|^{2}\Big(\be\Big[\Big|\int_0^T\big|Z^\alpha(s)|^2ds\big|^{ap}\Big]\Big)^{\frac{1}{ap}},
\eeqo
where $G$ is a value of the moment generating function for a normally distributed random variable which is continuous in $\beta$. We remark that $\be[|\int_0^T|Z^\alpha(s)|^2ds|^{ap}]<\infty$ holds for a sufficiently small $ap>1$, which could be proved as in \eqref{squarezz}. The limit $\lim_{\beta\rightarrow\alpha}\be\big[|X^\alpha(t)-X^\beta(t)|^2\big]=0$ is proved. The limit $\lim_{\beta\rightarrow\alpha}\be\big[\int_0^T|Z^\alpha(t)-Z^\beta(t)|^2dt\big]=0$ is proved analogously. The continuity $\beta\mapsto\rho_t^\beta(\xi)$ follows.\\
\noindent We calculate the limits. The limit $\lim_{\beta\rightarrow 0}\rho_{t}(\xi)=\be[\xi|\bfi_t]$ can be deduced from continuity.
We deal with the limit $\beta\rightarrow\infty$. From \eqref{twofactors} we derive the bounds
\beqo
\underline{M}^\beta(t)=X^\beta(t)-\beta\int_0^t\ln(\frac{T}{s})Z^\beta(s)ds\leq Y^\beta(t)\leq X^\beta(t)=\overline{M}^\beta(t),\quad 0\leq t\leq T.
\eeqo
For each $t\in[0,T]$ the mapping $\beta\mapsto X^\beta(t)$ is non-increasing and uniformly bounded from below, and we conclude that $\lim_{\beta\rightarrow\infty}X^\beta(t)=\inf_{\beta>0}\be^{\bqq^\beta}[\xi|\bfi_t]=-||-\xi||^\infty_t$ a.s.. As we assume that $\be[\sup_{t\in[0,T]}\big| ||-\xi||_t^\infty\big|^2]<\infty$ we can prove by dominated convergence theorem  that $X^\beta(t)$ converges strongly in $\mathbb{L}^2(\Omega, \bfi_t,\bp;\mathbb{R})$ and weakly in $\mathbb{L}^2(\Omega,\bfi_t,\bp;\mathbb{R})$ for any $0\leq t\leq T$. To fit our dynamics into the framework of Peng \cite{penglimit} let us consider the process $\tilde{X}^\beta(t)=-X^\beta(t)$. We obtain
\beqo
\tilde{X}^\beta(t)&=&-\xi+\int_t^T\beta\ln(\frac{T}{s})Z^\beta(s)ds+\int_t^TZ^\beta(s)dW(s),\\
&=&\tilde{X}^\beta(0)-A^\beta(t)-\int_0^tZ^\beta(s)dW(s),\quad 0\leq t\leq T,
\eeqo
with the non-decreasing continuous process $A^\beta(t)=\int_0^t\beta\ln(\frac{T}{s})Z^\beta(s)ds$, $0\leq t\leq T$ (recall that $Z\geq 0$). For each $t\in[0,T]$ the mapping $\beta\mapsto \tilde{X}^\beta(t)$ is non-decreasing and $\lim_{\beta\rightarrow\infty}\tilde{X}^\beta(t)=||-\xi||^\infty_t$, a.s..
Notice that $||-\xi||_t^\infty$ is a non-increasing process without a martingale part with the following representation
\beqo
||-\xi||_t^\infty=||-\xi||_0^\infty-\big(||-\xi||_{0}^\infty-||-\xi||_t^\infty\big)=||-\xi||_0^\infty-A(t),\quad 0\leq t\leq T,
\eeqo
with the non-decreasing process $A$. From Lemma 2.2 in Peng \cite{penglimit} we conclude that $||-\xi||_t^\infty$ is a right-continuous process. From Theorem 2.4 in Peng \cite{penglimit} we next conclude that $A^\beta(t)$ converges weakly in $\mathbb{L}^2(\Omega, \bfi_t, \bp;\mathbb{R})$ to $A(t)$ for each $t\in[0,T]$. By combining the weak convergence of $A^\beta$ and $X^\beta$ we derive the weak convergence of $\underline{M}^\beta(t)$ to $-||-\xi||_0^\infty$ in $\mathbb{L}^2(\Omega, \bfi_t, \bp;\mathbb{R})$ for any $0\leq t\leq T$. \\
\noindent 6. Let $(Y^n, Z^n), (Y,Z)$ denote the solutions to \eqref{bsdez} under the terminal conditions $\xi^n, \xi$. Similarly as in \eqref{estimateyy} we obtain
\beqo
\lefteqn{\be^\bqq\Big[\Big(\int_0^T|Y^n(t)-Y(t)|^2\Big)^{1+\frac{\epsilon}{4}}\Big]}\\
&&\leq L\Big(\be^\bqq\Big[|\xi^n-\xi|^{2+\frac{\epsilon}{2}}\Big]+\be^\bqq\Big[\Big(\int_0^T|Z^n(t)-Z(t)|^2dt\Big)^{1+\frac{\epsilon}{4}}\Big]\Big).
\eeqo
Burkholder and Doob's inequalities, together with the martingale representation of $\xi^n-\xi$ yield the estimate
\beqo
\lefteqn{\be^\bqq\Big[\Big(\int_0^T|Z^n(t)-Z(t)|^2dt\Big)^{1+\frac{\epsilon}{4}}\Big]\leq L\be^\bqq\Big[\sup_{0\leq u\leq T}\Big|\int_0^u(Z^n(t)-Z(t))d\tilde{W}(t)\Big|^{2+\frac{\epsilon}{2}}\Big]}\\
&\leq& L\be^\bqq\Big[\Big|\int_0^T(Z^n(t)-Z(t))d\tilde{W}(t)\Big|^{2+\frac{\epsilon}{2}}\Big]\\
&=&L\be^\bqq\Big[\big|\xi^n-\be[\xi^n]-\xi+\be^\bqq[\xi]\big|^{2+\frac{\epsilon}{2}}\Big]\leq L\be^\bqq\big[|\xi^n-\xi|^{2+\frac{\epsilon}{2}}\big].\ \quad \ \quad \ \quad \
\eeqo
We can prove the convergence of $\be\big[\int_0^T|Y^n(t)-Y(t)|^2\big]\rightarrow 0$ by changing the measure to $\bqq$, as in \eqref{squarezz}, and applying the above estimates together with \eqref{integrabilitypower}.
\cbdu
\indent Some key properties of the static and dynamic operator are again different.\\
\noindent \textbf{Example 4.2:} The dynamic operator $\rho_t$ is not monotonic with respect to the terminal condition: $\xi\geq \eta$ may not imply $\rho_t(\xi)\geq \rho_t(\eta), \ 0<t< T$.\\
\noindent Take $\xi=e^{2W(T)-2 T}$. We have the representation
\beqo
\be^{\bqq}\big[e^{2W(T)-2T}|\bfi_t\big]&=&\be^{\bqq}\big[e^{2\tilde{W}(T)-2T-2\beta T}\big]\\
&=&e^{2\tilde{W}(t)-2t-2\beta T}=1+2\int_0^te^{2\tilde{W}(s)-2s-2\beta T}d\tilde{W}(s),\quad 0\leq t\leq T,
\eeqo
and by \eqref{solz} we can derive the solution to the BSDE \eqref{bsdez} in the form of
\beqo
Y(t)=e^{2\tilde{W}(t)-2t-2\beta T}-2\beta\ln(\frac{T}{t})\int_0^te^{2\tilde{W}(s)-2s-2\beta T}ds,\quad 0\leq t\leq T.
\eeqo
Fix $t\in(0,T)$. By recalling Theorem 2.1 from Matsumoto \& Yor \cite{yor} for joint continuous distribution of $\big(\int_0^{t}e^{2\tilde{W}(s)-\frac{1}{2}s}ds;\tilde{W}(t))$ on $(0,\infty)\times\mathbb{R}$ we conclude that
\beqo
\bqq\Big(\int_0^te^{2\tilde{W}(s)-2s}ds>Le^{2\tilde{W}(t)-2t}\Big)>0,
\eeqo
holds for an arbitrary constant $L>0$. The random variable $Y(t)$ can take negative values with a positive probability.
\cbdu
\indent Monotonicity of $\beta\mapsto\rho^\beta_t(\xi)$ and the lower bound from Proposition 4.2 cannot be extended as well into the dynamic setting.\\
\noindent \textbf{Example 4.3:} The dynamic operator $\rho_t$ is not conditionally invariant with respect to the terminal condition: $\rho_{s,T}(\xi)=\xi\rho_{s,T}(1), \ t\leq s\leq T, 0< t< T,$ for an $\bfi_t$-measurable $\xi$ may not hold.\\
Fix $t\in(0,T)$ and choose an $\bfi_t$-measurable, square integrable, non-constant random variable $\xi$. From \eqref{solz} we obtain
\beqo
\rho_{t,T}(\xi)=\xi-\beta\ln(\frac{T}{t})\int_0^tZ(s)ds<\xi=\xi\rho_{t,T}(1).
\eeqo
\cbdu

\section{Economic implications}
\indent Let us comment on economic implications of the valuations rules derived from the proposed generators. Our generators model a disappointment effect and a volatility aversion which cannot be obtained in the framework of classical BSDEs without delays. The time-delay plays an important role here as it allows a non-trivial feedback of the past experiences into the future expectation.\\
\indent The dynamic valuation rule based on the solution to our first BSDE \eqref{bsdey} can be represented as
\beq\label{priceyint}
\rho_{t,T}(\xi)&=&\psi(0,t,T)V(t)-\int_0^t(V(s)-V(t))\psi'(s,t,T)ds,\quad 0\leq t\leq T,\nonumber\\
V(t)&=&\be[\xi|\bfi_t],\quad 0\leq t\leq T,
\eeq
where $\psi'(s,t,T)>0$, see \eqref{derivativepsi}. Let us recall that Bell \cite{bell} postulates that the utility of a pay-off should be measured as
\beqo
Total\ utility=economic\ pay-off\ +\ psychological\ satisfaction,
\eeqo
where a psychological satisfaction is negative for a disappointment and positive for an elation, and the disappointment is proportional to the difference between what the investor expected and what he/she gets. We can conclude that our dynamic valuation rule \eqref{priceyint} fits perfectly into the valuation framework proposed in the disappointment model of Bell \cite{bell}. The integral term in \eqref{priceyint} models a psychological satisfaction which acts as a penalty/reward for not meeting/exceeding the expectations formed in the past. Recall Example 3.2 where we show that the integral can take a positive or negative value which could be clearly interpreted in this example as a disappointment or elation. Under \eqref{priceyint} if the pay-off $V(t)$ falls short of its prior expectations $(V(s), 0\leq s\leq t)$, then in addition to the value of the pay-off, the investor experiences some degree of disappointment; whereas if the pay-off is better than its prior expectations the investor feels some measure of elation. The greater the disparity between the expectations and the outcome, the greater disappointment/elation. The strength of the feedback of the past experiences is measured by the parameter $\beta$ which determines $\psi'(s,t,T)$. Our interpretations derived from \eqref{priceyint} coincide with the conclusions from the disappointment model of Loomes \& Sugden \cite{loomes} and agree with the behaviour incorporated in the disappointment model of Dybvig \& Rogers \cite{rogers} in which "the agents feel different if they are disappointed because they expected to do much better and feel lucky because they expected to do much worse". Moreover, as advocated in Rozen \cite{rozen}, a short-term increase in the value of $\xi$ does not have to imply an increase in the value $\rho_t$. Only if an increase in $\xi$ is long-term, in the sense that the current expectation $V(t)$ is increasing and dominates all the past expectations $(V(s), 0\leq s\leq t)$, then our price for $\xi$ increases. Such an effect seems to correspond to the gains monotonicity axiom from Rozen \cite{rozen}. Finally, notice that under \eqref{priceyint} high expectations make the investor "happy" now, but carry a significant cost if they are not met. Recall that the paid bonus of 5000\$ could be assessed as a loss relative to the expectation of 10000\$. If an increasing dynamics of $\xi$ (high hopes) is followed by a sharp downturn in $\xi$ (poor outcome), see Example 4.1, than a negative value $\rho_t<0$ arises which indicates an extreme disappointment of the investor. \\
\indent The dynamic valuation rule based on the solution to our second BSDE \eqref{bsdez} can be represented as
\beq\label{pricezint0}
\rho_{t,T}(\xi)&=&\tilde{V}(t)-\beta\ln(\frac{T}{t})\int_0^tZ(s)ds,\quad 0\leq t\leq T,\nonumber\\
\tilde{V}(t)&=&\be^{\bqq}[\xi|\bfi_t],\quad 0\leq t\leq T.
\eeq
We could model an aversion against the experienced volatilities. In \eqref{pricezint0} the integral term always acts as a penalty. Under the rule \eqref{pricezint0} the current expectation of the pay-off $\xi$ is penalized by the experienced volatilities in the past expectations. We can see that the current expectation of $\xi$ is heavily penalized if the investor experienced very high volatilities in the expectations of $\xi$ in the past. If an increasing trend in the volatilities of $\xi$ (an indicator of a riskier investment and a requirement for a higher return) is followed by a sharp downturn in $\xi$ (poor outcome), see Example 4.2, then a negative value $\rho_t<0$ arises which indicates an extreme "disappointment" of the investor. Notice that a time-delayed generator $\frac{1}{t}\int_0^tZ(s)ds$ penalized the valuations more strongly than the classical generator $Z(t)$ which could not lead to negative prices.\\
\indent An interesting insight into the valuation rule \eqref{pricezint0} could be gained from the following representation
\beq\label{pricezint}
\rho_{t,T}(\xi)&=&\tilde{V}(0)+\int_0^tdR^t(s),\quad 0\leq t\leq T,\nonumber\\
dR^t(s)&=&d\tilde{V}(s)-\beta\ln(\frac{T}{t})\sqrt{\frac{d[\tilde{V},\tilde{V}](s)}{ds}}ds,
\eeq
where $[,]$ denotes a quadratic variation process. We can deduce that the price of $\xi$ arises from the changes in the expectation of the pay-off $d\tilde{V}(s)$ penalized by the volatilities of these changes $\sqrt{\frac{d[\tilde{V},\tilde{V}](s)}{ds}}ds$. Under \eqref{pricezint} the investor requires a compensation for the volatility risk he/she is taking. The value of $\xi$ increases only if the realized gain in the expected pay-off exceeds the volatility. If the required return is not met, a disappointment effect arises. The rule \eqref{pricezint} could model preferences of an investor who follows Sharpe ratio criterion as the return is compared to the volatility.  \\
\indent Our two models based on BSDEs with time-delayed generators provide decision rules which are consistent with the behaviours observed in the economic literature.

\section{Conclusion}
\indent BSDEs with time-delayed generators are a new research area both from the point of view of the theory and applications. We hope that these equations could offer a possibility of taking into account non-trivial dynamic investors' behaviours and preferences, such as non-monotone preferences. Further mathematical study of these equations is desirable.

\appendix
\section{Some results on modified Bessel functions}
For the reader's convenience we give some properties of the modified Bessel functions of the first and second kind which we use in Section 3. These properties can be found in Chapter 9.6 in Abramowitz and Stegun \cite{abramowitz}. We are only interested in $w\mapsto I_0(w), w\mapsto I_1(w), w\mapsto K_0(w), w\mapsto K_1(w)$ which are defined on the non-negative real axis $[0,\infty)$.\\
\indent The functions $w\mapsto I_0(w), w\mapsto I_1(w)$ are continuous, non-negative and strictly increasing, with the limits
\beq\label{limiti}
I_\nu(w)&\sim&\big(\frac{1}{2}w\big)^\nu,\quad w\rightarrow 0,\quad \nu=0,1,\nonumber\\
I_\nu(w)&\sim&\frac{e^w}{\sqrt{2\pi w}},\quad w\rightarrow\infty,\quad \nu=0,1.
\eeq
The following relations hold
\beq\label{derivativei}
I'_1(w)=I_{0}(w)-\frac{1}{w}I_1(w), \quad I'_0(w)=I_1(w),
\eeq
from which we can derive
\beq\label{integrationi}
\int w I_1(w)dw=wI_0(w),\quad \int w I_0(w)dw=wI_1(w).
\eeq
\noindent The functions $w\mapsto K_0(w), w\mapsto K_1(w)$ are continuous, non-negative and strictly decreasing, with the limits
\beq\label{limiki}
K_0(w)&\sim&-\ln w, \quad K_1(w)\sim w^{-1},\quad w\rightarrow 0,\nonumber\\
K_\nu(w)&\sim&e^{-w}\sqrt{\frac{\pi}{2w}},\quad w\rightarrow\infty,\quad \nu=0,1.
\eeq
The following relations hold
\beq\label{derivativek}
-K'_1(w)=K_{0}(w)+\frac{1}{w}K_1(w), \quad K'_0(w)=-K_1(w)
\eeq
from which we can derive
\beq\label{integrationk}
\int w K_1(w)dw=-wK_0(w),\quad \int w K_0(w)dw=-wK_1(w).
\eeq
\noindent The key relation between $I$ and $K$ is
\beq\label{wronskian}
I_0(w)K_1(w)+I_1(w)K_0(w)=\frac{1}{w}.
\eeq


\begin{thebibliography}{}
\bibitem{abramowitz}
Abramowitz, M.; Stegun, I.A. \emph{Handbook of Mathematical Functions with Formulas, Graphs, and Mathematical Tables}; Dover Publications Inc., 1965.
\bibitem{antonelli}
Antonelli, F.E.; Barucci, E.; Mancino, M. Asset pricing with a forward-backward stochastic differential utility. Economic Letters \textbf{2001}, 72, 151-157.
\bibitem{barrieu}
Barrieu, P.; El Karoui, N. Pricing, hedging and optimally designing derivatives via minimization of risk measure. In \emph{Indifference Pricing}; Carmona, R., Ed.; Princeton University Press, 2008; 77-141.
\bibitem{bell}
Bell, D. Disappointment in decision making under uncertainty. Operations Research \textbf{1985}, 33, 1-27.
\bibitem{delong1}
Delong, {\L}.; Imkeller, P. Backward stochastic differential
equations with time delayed generators - results and
counterexamples. The Annals of Applied Probability \textbf{2010}, 20, 1512-1536.
\bibitem{delong2}
Delong, {\L}.; Imkeller, P. On Malliavin's differentiability of BSDEs with time-delayed generators driven by a Brownian motion and a Poisson random measure. Stochastic Processes and their Applications \textbf{2010}, 9, 1748-1775.
\bibitem{detemple1}
Detemple, J.; Zapatero, Z. Asset prices in exchange economy with habit formation. Econometrica \textbf{1991}, 59, 1633-1657.
\bibitem{detemple}
Detemple, J.; Zapatero, Z. Optimal consumption portfolio with habit formation. Mathematical Finance \textbf{1992}, 2, 251-274.
\bibitem{gonzalo}
Dos Reis, G.; Reveillac, A.; Zhang, J. FBSDE with time delayed generators - $L^p$ solutions, differentiability,
representation formulas and path regularity. Preprint \textbf{2010}.
\bibitem{duffie}
Duffie, D.; Epstein, L. Stochastic differential utility. Econometrica \textbf{1992}, 60, 353-394.
\bibitem{rogers}
Dybvig, P.; Rogers, L. C. G. High hopes and disappointment. Preprint \textbf{2010}.
\bibitem{K}
El Karoui, N.; Peng, S.; Quenez, M.C. Backward stochastic
differential equations in finance. Mathematical Finance \textbf{1997}, 7, 1-71.
\bibitem{ravanelli}
El Karoui, N.; Ravanelli, C. Cash sub-additive risk measures and interest rate ambiguity. Preprint \textbf{2008}.
\bibitem{koszegi}
K\H{o}szegi, B.; Rabin, M. A model of reference-dependent preferences. The Quarterly Journal of Economics \textbf{2006}, 121, 1133-1165.
\bibitem{lazrak}
Lazrak, A.; Quenez, M.C. A generalized stochastic differential utility. Mathematics of Operations Research \textbf{2003}, 28, 154-180.
\bibitem{loewen}
Loewenstein, G. Anticipation and the valuation of delayed consumption. The Economic Journal \textbf{1987}, 97, 666-684
\bibitem{loewen2}
Loewenstein, G.; Prelec, D. Preferences for sequences of out-comes. Psychological Review \textbf{1993}, 100, 91.108.
\bibitem{loomes}
Loomes, G.; Sugden, R. Disappointment and dynamic consistency under uncertainty. The Review of Economic Studies \textbf{1986}, 53, 271-282.
\bibitem{yor}
Matsumoto, H.; Yor, M. Exponential functionals of Brownian motion: probability laws at fixed time. Probability Survey \textbf{2005}, 2, 312-347.
\bibitem{gianin}
Rosazza Gianin, E. Risk measures via $g$-expectations. Insurance: Mathematics and Economics \textbf{2006}, 39, 19-34.
\bibitem{penglimit}
Peng, S. Monotonic limit theorem of BSDE and nonlinear decomposition theorem of Doob-Meyer's type. Probability Theory and Related Fields \textbf{1994}, 113, 473-499.
\bibitem{pol}
Polyanin, A.; Zaitsev, V. \emph{Handbook of Exact Solutions for Ordinary Differential Equations}; Chapman\&Hall CRC, 2003.
\bibitem{protter}
Protter, P. \emph{Stochastic Integration and Differential Equations}; Springer-Verlag, 2004.
\bibitem{rozen}
Rozen, K. Foundations of intrinsic habit formation. Econometrica \textbf{2010}, 78, 1341-1373.

\end{thebibliography}
\end{document}